\documentclass[12pt,english,a4paper]{article}

\pdfoutput=1

\usepackage{amsmath,amsfonts,amssymb,babel,slashed,color,empheq,tensor}
\usepackage{wasysym}
\usepackage{jheparxiv}
\usepackage[utf8]{inputenc}
\usepackage{amsmath}
\usepackage{amsfonts}
\usepackage{amssymb}
\usepackage{latexsym}
\usepackage{mathrsfs}
\usepackage{braket}		
\usepackage{setspace}
\usepackage{mathrsfs}
\usepackage{graphicx}
\usepackage{skak}
\usepackage{color}
\usepackage{xcolor}
\usepackage{mathtools}
\usepackage{slashed}
\usepackage{twistor}
\usepackage[all]{xy}
\usepackage{tikz-cd}
\usepackage{slashed}

\usepackage{empheq}
\usepackage{float}
\usepackage{mathtools}
\usepackage{tikz}
\usepackage{roundrule}
\usepackage{stackengine}
\usepackage{scalerel}
\def\rlwd{.9pt}
\def\lhexbrace{\kern1pt%
\setstackgap{S}{0pt}\def\stackalignment{l}
\ThisStyle{\scalerel*{%
  \stackunder[-\rlwd]{%
    \stackon[-\rlwd]{\roundrule{\rlwd}{4pt}}{\rotatebox{60}{\roundrule{4pt}{\rlwd}}}%
  }{\rotatebox{-60}{\roundrule{4pt}{\rlwd}}}%
}{\SavedStyle[}}}
\def\rhexbrace{%
\setstackgap{S}{0pt}\def\stackalignment{r}
\ThisStyle{\scalerel*{%
  \stackunder[-\rlwd]{%
    \stackon[-\rlwd]{\roundrule{\rlwd}{4pt}}{\rotatebox{-60}{\roundrule{4pt}{\rlwd}}}%
  }{\rotatebox{60}{\roundrule{4pt}{\rlwd}}}%
}{\SavedStyle[}}\kern1pt}

\makeatletter

\newcommand{\bep}{\begin{picture}}
\newcommand{\eep}{\end{picture}}

\newcounter{YoungHeight}\newcounter{YoungWidth}

\newcounter{Mul1}\newcounter{Mul2}\newcounter{Mul3}\newcounter{Mul4}
\newcounter{A0}\newcounter{A1}\newcounter{A2}
\newcounter{B3}
\newcounter{C3}\newcounter{C4}
\newcounter{D1}\newcounter{D2}\newcounter{D3}
\newcounter{T0}\newcounter{T1}

\newlength{\txtHShift}

\newlength{\txtWidth}

\newcommand{\tG}{\mathtt{G}}

\newcommand{\kin}{\mathrm{kin}}

\newcommand{\ib}{\boldsymbol{\mathtt{I}}}
\newcommand{\jb}{\boldsymbol{\mathtt{J}}}
\newcommand{\kb}{\boldsymbol{\mathtt{K}}}
\newcommand{\lb}{\boldsymbol{\mathtt{L}}}

\newcommand{\Mat}{\mathrm{End}}
\newcommand{\YM}{\text{YM}}

\newcommand{\sX}{\boldsymbol{\mathsf{X}}}

\newcommand{\ttb}{\underline{\mathtt{t}}}

\newcommand{\Ibold}{\boldsymbol{I}}
\newcommand{\Jbold}{\boldsymbol{J}}
\newcommand{\Kbold}{\boldsymbol{K}}
\newcommand{\Lbold}{\boldsymbol{L}}

\newcommand{\sQ}{\mathsf{Q}}

\newcommand{\mso}{\mathfrak{so}}

\newcommand{\Tr}{\text{Tr}}
\newcommand{\cI}{\mathcal{I}}
\newcommand{\ta}{\mathtt{a}}

\newcommand{\boldJ}{\boldsymbol J}

\newcommand{\tp}{\mathsf{p}}

\newcommand{\UV}{\mathrm{UV}}

\usepackage{tcolorbox}
\usepackage{graphicx}

\newcommand{\nn}{\nonumber}

 \def\one{\mbox{1 \kern-.59em {\rm l}}}

\newcommand{\cK}{\mathcal{K}}

\newcommand{\m}{\mathrm{m}}

\newcommand{\V}{\, V}

\newcommand{\eps}{\epsilon}

\newcommand{\g}{\mathfrak{g}}
\newcommand{\msf}[1]{\mathsf{#1}}

\newcommand{\msu}{\mathfrak{su}}
\def\mmu{\mathfrak{u}}

\newcommand{\mg}{\mathfrak{g}}
\newcommand{\mk}{\mathfrak{y}}

\newcommand{\mb}{\mathtt{M}}
\newcommand{\nb}{\mathtt{N}}

\newcommand{\oneloop}{\text{1-loop}}
\newcommand{\tJ}{\mathtt{J}}

\def\a{\alpha}  \def\b{\beta}

\def\l{\lambda} \def\L{\Lambda}  

\makeatother

\begin{document}

 \begin{flushright}
  UWThPh-2024-11 
 \end{flushright}
 \vspace{-10mm}
 
\title{\Large Quantum $\hs$-Yang-Mills from the IKKT matrix model}

\author[]{Harold C. Steinacker}
\affiliation[]{Department of Physics, University of Vienna, \\
Boltzmanngasse 5, A-1090 Vienna, Austria}
\emailAdd{harold.steinacker@univie.ac.at}

\author[]{\& Tung Tran}

\emailAdd{tung.tran@univie.ac.at}

\abstract{We study the one-loop effective action of the higher-spin gauge theory induced by the IKKT matrix model on a $\cM^{1,3}\times \cK$ background, where  
$\cM^{1,3}$ is an FLRW cosmological spacetime brane and $\cK$ 
are compact fuzzy extra dimensions. 
In particular, we show that all non-abelian ($\hs$-valued) gauge fields in this model acquire mass via quantum effects, thus avoiding no-go theorems. This leads to a massive non-abelian quantum $\hs$-Yang-Mills theory
, whose detailed structure
depends on $\cK$. 
The stabilization of $\cK$ at one loop is understood as a result of the coupling between $\cK$ and the $U(1)$-flux bundle on space-time. This flux stabilization induces the KK scale into the $\cN=4$ SYM sector of the model, which break superconformal symmetry.}

 \maketitle

\section{Introduction}

The present paper is part of a larger program aiming to obtain near-realistic low-energy physics from the IKKT matrix model. The IKKT matrix model \cite{Ishibashi:1996xs} may be viewed as a constructive non-standard description of type IIB superstring theory, and it offers a novel approach to connect physics at the Planck scale to our observable world described by the Standard Model and general relativity. 

At the classical level, the IKKT matrix model can be viewed as a ``pre-gravity'' model for a space-time brane $\cM^{1,3}\xhookrightarrow{}\R^{1,9}$  embedded in target space, in a suitable decoupling regime\footnote{See e.g. \cite{Steinacker:2019fcb,Steinacker:2024unq} for a review.  Note that this approach is completely different from other approaches to matrix models 
which focus on the physics in the bulk (rather than on the brane), see e.g. \cite{Anous:2019rqb}.}. 
A (higher-spin) extended variant of GR then arises on $\cM^{1,3}$ at the quantum level \cite{Steinacker:2021yxt,Steinacker:2023myp}, in the spirit of induced gravity \cite{Sakharov:1967pk}. This quantum theory is UV finite due to the maximal supersymmetry of the model, but requires the presence of compact fuzzy extra dimensions $\cK \subset\R^6$ in the transversal directions with finitely many dof. per unit volume. 
The matrix fluctuations on such a background can be interpreted in terms of open strings propagating on the brane. For some pertinent 
recent development in this direction see e.g. \cite{Battista:2023glw,Kumar:2023bxg}, and 
\cite{Kim:2011cr,Anagnostopoulos:2022dak,Nishimura:2019qal} for numerical investigation of the emergence of space-time  within this model.

To avoid explicit Lorentz-violating $B$-fields on space-time, we consider a manifestly $SO(1,3)$-invariant  FLRW cosmological space-time brane $\cM^{1,3}$ with a twisted $S^2_{\tJ}$-bundle structure \cite{Steinacker:2017vqw}. 
Fluctuations on such covariant quantum spaces give rise to a spectrum of higher-spin ($\hs$) valued fields on space-time.
In particular, this leads to $\hs$-valued Yang-Mills gauge fields arising on stacks of coinciding branes, generalizing the familiar mechanism from non-commutative gauge theory and string theory \cite{Aoki:1999vr,Madore:2000en}. This higher-spin gauge theory is free of ghosts (negative norm states) and is almost-local \cite{Sperling:2019xar,Steinacker:2019awe}.\footnote{The almost-commutative $U(1)$-sector with only gravitational interactions has been shown to bypass Weinberg's soft theorem and have non-trivial $S$-matrices in the flat and late-time regime, which, nevertheless, are exponentially suppressed by the time-dependent couplings \cite{Steinacker:2023cuf,Steinacker:2023ntw}.} It is also quite different from conventional $4d$ local higher-spin theories such as (quasi-) chiral theories \cite{Metsaev:1991mt,Metsaev:1991nb,Ponomarev:2016lrm,Ponomarev:2017nrr,Skvortsov:2018jea,Metsaev:2018xip,Metsaev:2019aig,Tsulaia:2022csz,Sharapov:2022faa,Sharapov:2022awp} and conformal higher-spin gravity \cite{Tseytlin:2002gz,Segal:2002gd,Bekaert:2010ky} which are non-unitary.


The purpose of this paper is to study the $\hs$-
YM theory arising from the IKKT model at one loop, and see if there is a mechanism which could make it 
sufficiently similar to the ordinary Yang-Mills gauge theory. Unfortunately, our finding is \emph{not} completely satisfactory. In particular, while we can show that all non-abelian ($\hs$-valued) gauge fields of the model acquire mass from quantum effects (thus avoiding no-go theorems), we are 
\emph{not} able to find a mechanism for creating a big mass gap between the lowest-spin sector and the higher-spin one. This means that higher-spin fields do \emph{not} decouple from the system in the low-energy limit, which, nonetheless, agrees with the common wisdom in conventional higher-spin theories. Note that our analysis requires extra dimensions $\cK$ in the transversal directions as well as
the $S^2_{\tJ}$-bundle over the space-time brane $\cM^{1,3}$. 

Our main results on the one-loop effective action for the non-abelian $\hs$-YM sector of the IKKT matrix model on $\cM^{1,3}$ can be summarized as follows:
\begin{itemize}
    \item[1-] 
    We first compute the one-loop effective action of the YM theory, which is UV finite and depends on the structure of $\cK$. The contributions of UV modes running in the loop is evaluated  using a geometric trace formula for the space of operators on $\cK$. 

     \item[2-] 
     We show that whenever $\cK$ is non-trivial and reducible,
     all non-abelian ($\hs$-valued) gauge fields will acquire quantum contributions to their mass, even in the unbroken sector. The mass scale is set by the Kaluza-Klein scale $m_{\cK}$. 
     In particular, the massive $\hs$-YM induced by the IKKT matrix model is not in conflict with standard no-go theorems in the low-energy regime. 
    
     \item[3-] We derive geometric formulas for one-loop effective potential 
     between separable sub-branes $\cK_i\subset \cK$. We show that attractive interactions  
     between separable branes arise for suitable brane configurations. In particular, self-intersecting brane structures are argued to be energetically favored.

 \item[4-] 

We compute the effective potential for $\cK$ as a function of $m_{\cK}$, 
resulting from the coupling between $\cK$ and the
$U(1)$-flux bundle on  $\cM^{1,3}$ at one loop. 
This coupling introduces a scale into the $\cN=4$ SYM sector within $\hs$-YM 
allowing the formation of non-trivial vacua in the scalar sector, which in turn determines the extra dimensions. 
We show that this potential exhibits a non-trivial minimum, thus stabilizing  
$\cK$. 
 This mechanism provides a crucial difference between  ordinary $\cN=4$ SYM and
 the low-energy $\cN=4$ SYM sector of $\hs$-YM, 
where the conformal symmetry is broken.
     
\end{itemize}

The paper is organized as follows. Section \ref{sec:2} sets the stage to study the emergent quantum non-abelian massive higher-spin gauge theory called $\hs$-YM on $\cM^{1,3}$  induced by the IKKT matrix model. Section \ref{sec:3} obtains the classical action for $\hs$-YM. Section \ref{sec:4} derives a general formula for the one-loop effective action on the background brane $\cM^{1,3}\times S^2_{\tJ}
\times \cK$. 
Using this formula, we study the one-loop effective action and compute the renormalization of the coupling constant of the quantum $\hs$-YM in Section \ref{sec:7}. In Section \ref{sec:massive}, we show that all (higher-spin) Yang-Mills fields acquire significant quantum corrections to their mass.
Finally, we compute the effective potential between separable branes $\cK_i\subset\cK$ and discuss their stabilization in Section \ref{sec:5}. Further discussion can be found in Section \ref{sec:discussion}, and some technicalities are collected in Appendices \ref{app:A}.





\section{Review: the IKKT matrix model \& higher spin}\label{sec:2}
\subsection{The IKKT model and covariant quantum space-time}
\paragraph{The IKKT model.} The $SO(1,9)$-invariant action of the IKKT matrix model reads
\small
\begin{align}\label{SO(1,9)action}
    S=\frac{1}{g^2}\Tr\Big([T^{\Ibold},T^{\Jbold}][T_{\Ibold},T_{\Jbold}]
    - \overline{\Psi}_{\boldsymbol{\cA}}(\Gamma^{\Ibold})^{\boldsymbol{\cA}}{}_{\boldsymbol{\cB}}[T_{\Ibold},\Psi^{\boldsymbol{\cB}}]\Big)\,, \qquad \Ibold=0,1,\ldots,9\,.
\end{align}
\normalsize
Here, $T^{\Ibold}$ are hermitian matrices acting on a Hilbert space $\cH$, and $\Psi$'s are matrix-valued Majorana-Weyl spinors of $\mso(1,9)$ with $\Gamma^{\Ibold}$ being the generators of the associated Clifford algebra. Note that $g$ is the coupling constant of the matrix model, which will later be related to the physical coupling.
 The above action is invariant under gauge transformations $\delta T^{\Ibold}=[T^{\Ibold},\xi]$ 
 and $\delta \Psi^{\boldsymbol{\cA}}=[\Psi^{\boldsymbol{\cA}},\xi]$
 for any $\xi\in \End(\cH)$. In order to introduce a scale and to break supersymmetry, it is convenient to  add a mass term to the original action \eqref{SO(1,9)action} of the IKKT matrix model. This allows to obtain  
non-trivial background solutions of the classical equations of motion such as the $SO(1,3)$-invariant FLRW cosmological spacetime cf. \cite{Steinacker:2017bhb} without taking into account quantum corrections.

\paragraph{Background brane  $\cM^{1,3} \times S^2_{\tJ}\times \cK$ and semi-classical description.} In the following, we shall consider a 
matrix background configuration, which describes a
``brane'' with the product structure $\cM^{1,3}\times S^2_{\tJ}\times \cK$. Here, $\cM^{1,3}$ models our FLRW cosmological spacetime, $S^2_{\tJ}$ is a internal fuzzy 2-sphere encoding higher-spin structures with cutoff at $\tJ+1$, and $\cK$ describes the fuzzy extra dimensions. 
Such a background is given by $9+1$ hermitian matrices 
\begin{align}\label{eq:background}
\underline{T}^{\Ibold}=\binom{\underline{T}^{\dot\mu}}{\underline{T}^{\ib}}\,,\qquad \dot\mu=0,1,2,3\,,\quad \ib=4,\ldots,9\,,
\end{align}
where $\underline{T}^{\dot\mu}\in \End(\cH_{\cM})$ are the generators defining 
the $SO(1,3)$-invariant $3+1$ dimensional spacetime $\cM^{1,3}$
(which implicitly includes an $S^2_{\tJ}$ fiber with flux), and $\underline{T}^{\ib} \in \End(\cH_\cK)$ are  generators defining a  compact symplectic space $\cK$ 
embedded in the 6 transversal directions.
Accordingly, 
the  Hilbert space $\cH$ associated with the full brane configuration factorizes\footnote{strictly speaking, the factorization of $\cH_\tJ$ describing the internal sphere holds only locally.} as 
\begin{align}\label{eq:Hibert-space-factorization}
    \cH=\cH_{\cM}\otimes\cH_{\tJ}\otimes \cH_{\cK}\,,
\end{align}
where $\cH_{\tJ}$ is the Hilbert space associated to $S^2_{\tJ}$. Then, in the semi-classical limit where matrices are effectively commutative, the background matrices $\underline{T}$ can be  replaced by semi-classical functions on $\cM^{1,3}\times \cK$ via the dequantization maps
\begin{align}
    \End(\cH_{\cM}\otimes \cH_{\cK})&\mapsto C^{\infty}(\cM^{1,3})\times C^{\infty}(\cK)\nn\\
    F(T) &\mapsto   f(t)=\langle \zeta|F(T)|\zeta\rangle
\end{align}
where $|\zeta\rangle$ are some localized quasi-coherent states. This amounts to a replacement
\begin{align}\label{eq:semi-background}
\underline{T}^{\Ibold}=\binom{\underline{T}^{\dot\mu}}{\underline{T}^{\ib}}\mapsto\ttb^{\Ibold}=\binom{\ttb^{\dot\mu}}{\ttb^{\ib}}\,,\qquad \qquad [.\,,.] \to \im\{.,.\}
\end{align}
where $\{\,,\}$ denote the Poisson brackets, whenever appropriate\footnote{To simplify the notation, this replacement will not always be spelled out.}.
Here, $\cM^{3,1}\times S^2_{\tJ}$, which is generated by the $\ttb^{\dot\mu}$, can be understood in terms of quantized twistor space. 
More precisely, the spacetime brane $\cM^{1,3}$ given by the covariant quantum space is known to carry a twisted $S^2_\tJ$-bundle structure  \cite{Sperling:2019xar}. 
In the semi-classical limit, one can choose
the following Cartesian and hyperbolic coordinate functions on $\cM^{3,1}$
\begin{align}
    \begin{pmatrix}
     y^0\\
     y^1\\
     y^2\\
     y^3
    \end{pmatrix}=R\cosh(\tau)\begin{pmatrix}
     \cosh(\chi)\\
     \sinh(\chi)\sin(\theta)\cos(\varphi)\\
     \sinh(\chi)\sin(\theta)\sin(\varphi)\\
     \sinh(\chi)\cos(\theta)
    \end{pmatrix}\, .
\end{align}
Here, $\tau$ a time-like parameter and $R$ is the curvature radius. 
Note that $\cM^{1,3}$ is obtained by projecting out the coordinate $y^4=R\sinh (\tau)$ of an underlying 4-hyperboloid $H^4$ with radius $R$ characterized by 4+1 embedding functions $y^a:\ \cM^{1,3} \hookrightarrow  \R^{1,4}$ transforming as vectors of $SO(1,4)$ where
\begin{align}
  \eta^{ab}y_ay_b=-R^2\,,\qquad \eta^{ab}=\diag(-,+,+,+,+)\,,
\end{align}
for $a=0,1,2,3,4$. 
The relation between the matrices and their semi-classical functions can be made explicit in terms of (quasi-) coherent states. In particular, 
\begin{align}
      y^a:=\langle \zeta|Y^a| \zeta\rangle\in \R^{1,4}\,,
\end{align}
where $Y^a$ are  hermitian matrices acting on $\cH_\cM$ and $|\zeta\rangle$ are some (quasi-) coherent states. 
A more systematic discussion can be found e.g. in  \cite{Steinacker:2019fcb,Steinacker:2024unq}.

Similarly, a geometric description 
for $\cK$ is obtained in terms of  coordinate functions $m_\cK z^{\ib}= \langle \zeta|
\ttb^{\ib}|\zeta\rangle$ for $\ib=4,\ldots,9$, where the $z^{\ib}$ are dimensionless. Here, $m_{\cK}$ is a Kaluza-Klein (KK) scale parameter, and
 such that $z^{\ib} z_{\ib} = \cO(1)$. In other words, the transversal generators $\ttb^{\ib}$ can be identified in the semi-classical regime with the classical embedding coordinates $z^{\ib}$ via
\begin{align}\label{t-z}
    \ttb^{\ib}\sim m_{\cK}\,z^{\ib} : \ \cK \hookrightarrow \R^6\,,\qquad \ib=4,\ldots,9\,.
\end{align}
These are the coordinates of some quantized compact symplectic space $\cK$, described by the finite-dimensional space of modes in $\End(\cH_{\cK})$ with 
\begin{align}\label{eq:dim-K}
    d_\cK := \dim \cH_\cK \ < \infty \ .
\end{align}
Note that there is a useful operator on $\cK$, from which we can extract the relevant Kaluza-Klein scales. This operator is the ``transversal" matrix Laplacian acting on $\End(\cH_\cK)$:
\begin{align}
\label{Box-K-def}
\Box_6 := [\ttb^{\ib},[\ttb_{\ib},.]]\,,
\end{align}
which defines the KK mass arising from $\cK$ as the eigenvalues of some eigen modes $\Upsilon_{\Lambda}$, i.e.
\begin{align}
    \Box_6\Upsilon_{\L} = m^2_\L\, \Upsilon_{\L} \ , \qquad m_{\L}^2 = m_\cK^2 \,\mu^2_{\L}\,. \ 
 \label{Box-6-K-EV}
\end{align}
Here, $\mu^2_{\L}$ characterizes the spectral geometry of $\cK$.
It will be essential  that
$d_{\cK}=\dim\cH_\cK$ is finite, cf. \eqref{eq:dim-K}. Then the number of associated KK modes induced by $\cK$ is also finite\footnote{This is crucial for obtaining finite KK mass from the extra dimensions $\cK$.}.

\paragraph{$\hs$-valued fluctuations.} 

Now consider fluctuations 
$\cA^{\Ibold}$ around the above background $\ttb^{\Ibold}$:
\begin{align}\label{eq:fluctuations}
    t^{\Ibold}=\ttb^{\Ibold}+\cA^{\Ibold}
    =\binom{\ttb^{\dot\mu}\otimes \one_\cK}{\one_{\cM^{1,3}}\otimes\ttb^{\ib}}
    +  \cA^{\Ibold}\,.
\end{align}
Taking into account the above geometric description of the background brane $\cM^{1,3}\times \cK$, the fluctuations $\cA^{\Ibold}$ in \eqref{eq:fluctuations} can be written as
\begin{align}
    \cA^{\Ibold} =   \binom{\cA^{\dot\mu}_{\a}(y)}{\phi^{\ib}_{\a}(y)}\l^{\a}\,,\qquad \dot\mu= 0,1, 2, 3\,,\qquad \ib=4,\ldots,9\,,\quad \alpha=1,\ldots,\dim(\mg)\,.
\end{align}
Here, $\cA^{\dot\mu}_{\a}(y)$ and $\phi^{\ib}_{\a}(y)$ are 4-dimensional $\mg\otimes\hs$-valued functions on $\cM^{1,3}$ where
\begin{align}
    \mg := \End(\cH_\cK)\,
\end{align}
with generators $\l^{\a}$ for $\alpha=1,\ldots,\dim(\mg)$. Similarly, the Majorana-Weyl spinors $\Psi^{\cA}$ \eqref{SO(1,9)action} also take values in $\mg\otimes \hs$, and can be written in a suitable way by decomposing the $10d$ gamma matrices as
\begin{align}\label{eq:Gamma-decomposition}
    \Gamma^{\Ibold}=(\Gamma^{\dot\mu},\Gamma^{\ib})\,,\qquad \Gamma^{\dot\mu}=\gamma^{\dot\mu}\otimes \one_8\,,\quad \Gamma^{\ib}=\gamma_5\otimes \gamma^{\ib}\,,
\end{align}
where $\gamma_5=\gamma_0\gamma_1\gamma_2\gamma_3=-\gamma_5^{\dagger}$ is the $4d$ chirality operator and $\gamma^{\ib}$ generate the $6d$ Euclidean Clifford algebra. As a result,  
\begin{align}\label{eq:fermion-splitting}
    \Psi_{\boldsymbol{\cA}}\mapsto\Psi_{A\cI}=(\Psi_{\ta \cI},\Psi_{\dot\ta\cI})\,,\qquad \ta=0,1\,,\quad \dot\ta=\dot 0, \dot 1\,,\quad \cI=1,2,3,4\,,
\end{align}
with $(\ta,\dot\ta)$ are $SO(1,3)$-spinor indices and $(\cI,\cJ)$ are indices in the bi-vector representations of $SU(4)$. 


In cooperating the $S^2_\tJ$ factor of $\cM^{1,3}$ into the above picture, the fluctuations $\cA$ become functions on $\cM^{1,3} \times S^2_{\tJ}$. Because this bundle is equivariant\footnote{i.e. local rotations on $\cM^{1,3}$ act non-trivially on the fiber $S^2_{\tJ}$.}, the harmonics on the internal $S^2_{\tJ}$ behave as higher-spin modes on spacetime. As such, these higher-spin modes 
can be realized by spherical harmonics  of $S^2_{\tJ}$, i.e.\footnote{This is similar to the standard realization of massless fields in 3$d$ higher-spin gravities (HSGRAs) \cite{Blencowe:1988gj,Bergshoeff:1989ns,Pope:1989vj,Fradkin:1989xt,Campoleoni:2010zq,Henneaux:2010xg,Gaberdiel:2010pz,Gaberdiel:2012uj,Gaberdiel:2014cha,Grigoriev:2019xmp,Grigoriev:2020lzu} using finite-dimensional higher-spin algebra.} 
\small
\begin{align}\label{eq:hs-valued-functions}
    \cA_{\dot\mu}^{\a}:=\sum_s\cA_{\dot\mu;sm}^{\a}(y)\hat Y^{sm}\,,\qquad  \phi_{\ib}^{\a}:=\sum_s\varphi^{\alpha}_{\ib;sm}(y)\hat Y^{sm}\,,\qquad \Psi_{\ta\cI}^{\alpha}=\sum_s\Psi_{\ta\cI;sm}^{\alpha}(y)\hat Y^{sm}\,.
\end{align}
\normalsize
Then, the finite-dimensional $\hs$ algebra is given by
\begin{align}\label{eq:def-hs}
    \hs =\End(\cH_{\tJ})
\cong 
\msu(\tJ+1)\,.
\end{align}


\paragraph{Local frames of $\cM^{1,3}$.} The Poisson brackets between $\ttb^{\dot\mu}$ and $y^{\nu}$ define a frame
\begin{align}\label{eq:PoissonM13}
    E^{\dot\mu\nu} := \{\ttb^{\dot\mu},y^{\nu}\}&=\frac{\eta^{\dot\mu\nu}}{R}y^4=\eta^{\dot\mu\nu}\sinh(\tau)\,,\qquad \eta^{\dot\mu\nu}=\diag(-,+,+,+)\,,
\end{align}
for $\dot\mu=\dot 0,\dot 1,\dot 2,\dot 3$ and $\nu=0,1,2,3$. Using the frame $E$, one can obtain the \emph{effective} metric and the dilaton $\rho$ as \cite{Sperling:2019xar}
\begin{align}\label{eq:eff-G}
    G_{\mu\nu} = \sinh(\tau) \eta_{\mu\nu}=\rho^2\gamma_{\mu\nu} \,,\qquad \rho^2=\sinh^3(\tau)\,,
    \qquad \gamma_{\mu\nu}:=E^{\dot\kappa}{}_{\mu}E_{\dot\kappa\nu}=\rho^{-2}G_{\mu\nu}\,.
\end{align}
The \emph{auxiliary} metric $\gamma_{\mu\nu}$ is similar to the open string metric in string theory. Note that the effective metric describes a cosmological FLRW space-time \cite{Steinacker:2017vqw} where the dilaton $\rho$ evolves as
\begin{align}
\label{rho-a(t)}
    \rho^2=\sinh^3(\tau)\sim e^{3\tau} \ \propto \ a(t)^2 \ .
\end{align}
Here,  $a(t)$ is the FLRW scale parameter. The inverse of the above reads
\begin{align}
    G^{\mu\nu}=\rho^{-2}\gamma^{\mu\nu}\,.
\end{align}
 Further useful relations between $\ttb$'s and $y$'s are collected in Appendix \ref{app:A}.

\paragraph{Matrix field strength.} The most important quantity that we will work with is the $10d$ commutator, or equivalently the ``matrix field strength":
\begin{align}\label{eq:F-10d-full}
    \cF^{\Ibold\boldJ}:= 
      -\im[T^{\Ibold},T^{\boldJ}] \, 
      \sim \{t^{\Ibold},t^{\boldJ}\}
\end{align}
with dimension $[mass^2]$,
which encodes the Yang-Mills field strength as the lowest-spin sector on backgrounds given by non-commutative branes.
In the above $4+6$ splitting of the background, the $10d$ field strength decomposes as
\begin{align}\label{eq:F-10d}
    \cF^{\Ibold\boldJ}:= -\im[T^{\Ibold},T^{\boldJ}]
    =\begin{pmatrix}
        -\im[T^{\dot\mu},T^{\dot\nu}] & -\im[T^{\dot\mu},T^{\jb}]\\ -\im[T^{\ib},T^{\dot\mu}] & -\im [T^{\ib},T^{\jb}]
    \end{pmatrix}\equiv \begin{pmatrix}
        \cF^{\dot\mu\dot\nu} & \cF^{\dot\mu\jb}\\
        \cF^{\ib\dot\mu} & \cF^{\ib\jb}
    \end{pmatrix}\,.
\end{align}
We can ignore the spinorial analogs of the field strength, since we do not consider non-trivial fermion condensation.

\subsection{Relevant scales}
\label{sec:K-scales}

There are several scales arising on the proposed background, which 
play important roles in the one-loop effective action and the resulting effective field theory. 

\paragraph{Scales on $\cM^{1,3}\times S^2_{\tJ}$.} The background $\cM^{1,3}\times S^2_{\tJ}$ inherits two particular scales: (1) the
 non-commutativity scale of space-time $L_{\rm NC}=\frac{R}{\sqrt{\tJ}}\cosh^{\frac{1}{2}}(\tau)=\sqrt{\tJ}\,\ell_p\cosh^{\frac{1}{2}}(\tau)$; (2) IR curvature scale $L_{\text{H}}=R\cosh(\tau)$. 
 These scales are measured in terms of the Cartesian coordinates and \emph{not} in terms of the (effective) metric, cf. \eqref{eq:eff-G}. Any process with wavelength $\lambda$ where $L_{\rm NC}\ll \l \ll L_{\rm H}$ is said to belong to the low energy physics. In this regime, the detailed structure of $\cK$ turns out to be irrelevant. 
 We shall assume that the $\hs$ cutoff $\tJ \gg 1$ is a very large integer, and focus on the late-time regime $\cosh(\tau) \gg 1$. These two large numbers will lead to a large hierarchy of scales, as shown in Sections \ref{sec:7} and \ref{sec:5}.

\paragraph{Scales on $\cK$.}

The fuzzy extra dimensions $\cK$ also induce several relevant scales,
which will show up in the one loop computations. 
As usual, the UV cutoff scale $\L_{\UV}$ and the IR scale $\L_{\rm IR}$ of $\cK$ are set by the maximal 
and minimal eigenvalues of $\Box_6$ cf. \eqref{Box-6-K-EV}. Hence,  
the KK modes on $\cK$ have mass ranging from $\L_{\rm IR}$ all the way up to $\L_{\rm UV}\sim \cO(m_{\cK})$, i.e.
\begin{align}
    {\rm spec}\Box_6 \subset (\L_{\rm IR}^2, \L_{\rm UV}^2), \qquad \L_{\rm IR} \ll \Delta_\cK \ll \L_{\rm UV}\,.
\end{align}
Here, the non-commutativity scale 
$\Delta_\cK^2= \cO(\cF^{\ib\jb})$  of $\cK$
is set by the background symplectic structure on $\cK$, 
separating the semi-classical regime from the deep quantum regime.
It is typically given by 
the Bohr-Sommerfeld quantization condition
\begin{align}
\label{Bohr-Sommerfeld}
d_{\cK}\equiv \dim\cH_\cK =\int_\cK \frac{\Omega_{\cK}}{{(2\pi)^{\frac{|\cK|}{2}}}}\,,\qquad |\cK|:=\dim (\cK)\,,
\end{align}
where the total volume of $\cK$ can fit 
$d_\cK$ quantum cells with volume $(\Delta_{\cK})^{|\cK|}$. This is amounts to the following relations:
\begin{align}\label{dK-mK}
d_\cK \Delta_\cK^{|\cK|} \approx m_\cK^{|\cK|} \qquad \text{or} \qquad \Delta_{\cK}\approx\frac{1}{d_{\cK}^{1/|\cK|}}m_{\cK}\,.
\end{align}
between $\Delta_{\cK}$ and the UV cutoff scale $\Lambda_{\UV}\sim \cO(m_{\cK})$, which also plays the role of the curvature radius for $\cK$. For instance, if $\cK$ is the standard fuzzy 2-sphere $S^2_N$, cf. \cite{Madore:2000en}, then these scales are given by\footnote{It is crucial noting that $S^2_{\tJ}$ is drastically different with $S^2_N$ due to the fact that its generators are almost commutative even for large $\tJ$, cf. Appendix \ref{sec:S2J-structure}.}
\begin{align}
\label{scales-K}
    \L_{\rm UV} \sim m_\cK\,,\quad \Delta_\cK = \frac 1{\sqrt{N}} m_\cK\,,\quad \L_{\rm IR} \sim  \frac 1N m_\cK\,.
\end{align}
Note that all of these KK scales 
are expected to be in the UV regime from the space-time point of view.
For some special geometries such as self-intersecting branes cf. \cite{Sperling:2018hys}, $\cK$ may admit, in addition, certain {\em fermionic} (would-be) zero modes which are far below $\L_{\rm IR}$, as well as massless gauge bosons arising from point branes.



\subsection{Pertinent traces}
\label{sec:traces}

From \eqref{eq:Hibert-space-factorization}, we have
\begin{subequations}
    \label{End-factorization}
    \begin{align}
    \tr_{\cH_{\cM}\otimes \cH_{\tJ}\otimes\cH_{\cK}}&=\tr_{\cM}\times\tr_{\cH_{\tJ}}\times \tr_{\cH_{\cK}}\,,\\
    \Tr_{\End(\cH_{\cM})\otimes \End(\cH_{\tJ})\otimes \End(\cH_{\cK})}&=\Tr_{\End(\cH_{\cM})}\times \Tr_{\End(\cH_{\tJ})}\times \Tr_{\End(\cH_{\cK})}\,,
\end{align}
\end{subequations}
where due to the factorization of the background branes, we have
\begin{align}
    \End(\cH) = \End(\cH_{\cM}) \otimes \End(\cH_{\tJ})\otimes \End(\cH_\cK)\,.
\end{align}
 Here, we denote the trace over $\cH$ as $\tr_{\cH}$ while the trace over $\End(\cH)$ will be denoted by $\Tr_{\End(\cH)}$ since they are used in two different contexts. In particular, the former is used in classical regime while the latter is relevant in the quantum scenario, cf. \cite{Steinacker:2022kji,Steinacker:2023myp}
. Note that we will write $\Tr_{\End(\cH_{\cK})}\equiv\Tr_{\mg}$ and $\Tr_{\End(\cH_{\tJ})}\equiv\Tr_{\hs}$ for simplicity.\\

$\symknight$ \underline{Semi-classical traces.} First, let us discuss the trace over $\cH_\cM\otimes \cH_{\tJ}$. In the semi-classical limit, one can replace
\begin{align}\label{tr-replace}
    \tr_{\cM} \ \mapsto \ \int \limits_{\cM^{1,3}}\mho_0\qquad \text{where}\qquad \mho_0:=\frac{R}{ \ell_p^4 \,y^4}dy^0dy^1dy^2dy^3=
    \frac{\sqrt{|G|}}{\ell_p^4\,\rho^2}d^4y
\end{align}
is the symplectic measure on $\cM^{1,3}$. 
When including higher-spin fields into this picture, the above measure gets modified to a measure of an underlying non-compact twistor space $\P^{1,2}\simeq \cM^{1,3}\times S^2_{\tJ}$ as
\begin{align}\label{eq:overall-J}
    \mho= \varpi\, \mho_0=\varpi \,\frac{\sqrt{|G|}}{\ell_p^4\,\rho^2}d^4y\,.
\end{align}
Here $\varpi$ is the volume form of $S^2_{\tJ}$
normalized such that
\begin{align}
\label{S2-volume-form}
    \tr_{\cH_\tJ} \mapsto \int_{S^2_{\tJ}}\varpi\,, \qquad \quad
\int_{S^2_{\tJ}}\varpi=\tJ\, .
\end{align}
This gives us the overall $\tJ+1
\approx 
\tJ$ factor (for large cutoff $\tJ$) in \eqref{eq:overall-J} after taking the trace of suitable spherical harmonic modes on $S^2_{\tJ}$. 
Note that we will drop the notation $\sqrt{|G|}$ and simply write $\sqrt{G}$ for simplicity since no confusion can arise.\\


$\symknight$ \underline{Traces in the loops.} 
In the 
loop computations, we will need a totally different type of traces over the space of modes in $\End(\cH)$. Firstly, the trace $\Tr_{\End(\cH_{\cM})}$ can be  
evaluated
using the following type of semi-classical trace formulas \cite{Steinacker:2023myp}
\begin{align}\label{eq:Tr-End-M}
    \Tr_{\End(\cH_\cM)} \ \sim \ \int\limits_{\cM^{1,3}} d^4y\sqrt{G}\int \frac{d^4k}{(2\pi)^4\sqrt{G}}\,
\end{align}
evaluating on Gaussian wave packets or plane waves. This formula is applicable to UV-convergent integrals.
As always, $k^{\mu}$ are the characteristic wave vector (or the canonical momenta) associated with the coordinates $y^{\mu}$. 
This description is based on the underlying quantization map between $\End(\cH_\cM)$ and $C^{\infty}(\cM^{1,3})$.
The trace over $\hs$-modes will be discussed 
when concrete computations show up.

On the other hand, the 
traces over $\End(\cH_\cK)$ for the 
$\cK$ is novel and interesting. 
Assuming that  $\cK$ is reasonably large, one can use the following non-local geometric trace formula 
cf. \cite{Steinacker:2022kji,Steinacker:2023myp}:
\begin{align}
\Tr_{\End(\cH_{\cK})} \cO 
 = \int\limits_{\cK}\! \frac{\Omega_x \Omega_y}{(2\pi)^{|\cK|}}\,
  \left(^x_{y} \right| \cO \left|^x_{y} \right) \,,\qquad |\cK|:=\dim\cK\,, 
 \label{trace-coherent-End}
\end{align}
which holds even
in the UV regime. Here, $\cO$ is some operator acting on $\End(\cH_{\cK})$, and $\left|^x_{y} \right)\equiv |x\rangle\langle y|$ denote string modes on $\cK$.
If $\cK$ consists of several constituent
 branes $\cK_i\subset \cK$, then the off-diagonal string modes 
stretching between
$\cK_i$ and $\cK_j$ branes
will mediate their
interactions.
These data can be naturally generalized to stacks of different branes.
\begin{figure}[ht!]
    \centering
    \includegraphics[scale=0.42]{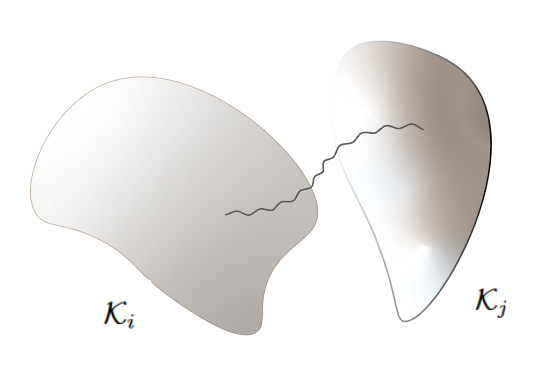}
    \caption{An open string stretching between two branes $\cK_i$ and $\cK_j$.}
    \label{fig:open-string-modes-KK}
\end{figure}

\section{Classical \texorpdfstring{$\hs$}{hs}-Yang-Mills theory 
}\label{sec:3}
This section derives the classical coupling constants of the non-abelian sector in the higher-spin gauge theory induced by the IKKT matrix model 
called $\hs$-YM.\footnote{Although this theory has a somewhat similar name to the one in \cite{Adamo:2022lah}, there should not be confusion between them. In particular, the HS-YM theory in \cite{Adamo:2022lah} is a quasi-chiral massless theory, while the $\hs$-YM in the present work is a massive one.} 
In the semi-classical limit, the full quantum bracket $[\,,]$ cf. \eqref{SO(1,9)action} decomposes as 
\begin{align}\label{eq:quantum-bracket}
    [a,b]=\im \{a,b\}+[a,b]_{\mg}
\end{align}
where $\{.,.\}$ denotes the Poisson bracket and $[.,.]_{\mg}$ denotes the Lie bracket of $\mg$. Using the frame \eqref{eq:PoissonM13}, we can rewrite the tangential fluctuations $\cA^{\dot\mu}$ as
\begin{align}
\label{A-frame}
    \cA^{\dot\mu} = E^{\dot\mu \mu} A_\mu \,,\qquad E^{\dot\mu\mu}=\{\ttb^{\dot\mu},y^{\mu}\}\,.
\end{align}
Then, in the effectively $4d$ regime, the tangential $\hs$-valued matrix field strength of Yang-Mills sector takes the form 
\begin{subequations}
    \begin{align}
    \cF^{\dot\mu\dot\nu} :&= -\im [\ttb^{\dot\mu}+\cA^{\dot\mu},\ttb^{\dot\nu}+\cA^{\dot\nu}]= E^{\dot\mu \mu} E^{\dot\nu \nu} \big(-\theta^{-1}_{\mu\nu} + F_{\mu\nu} \big) \,, \\
    F_{\mu\nu} :&= \del_\mu A_\nu -  \del_\nu A_\mu + [A_\mu,A_\nu]\,,
\end{align}
\end{subequations}
where we have dropped terms which encode derivatives $\del E$  of the frames, in the absence of strong gravity. The non-abelian (higher-spin) Yang-Mills gauge fields  read\footnote{Of course, one can use the standard polynomial representations for  $\hat Y$ as bookkeeping device for the internal higher-spin modes arising on $\cM^{1,3}$ as in the usual twistor construction for self-dual and chiral higher-spin theories, see e.g. \cite{Tran:2022tft,Herfray:2022prf,Adamo:2022lah}. However, the spherical harmonics $\hat Y$ allow us to keep the notations simple.}
\begin{align}
\label{gaugefields-l-hs}
    A_{\mu} = A_{\mu;\a sm}\l^\a\hat Y^{sm} \,\quad \in \End(\cH_{\cM})\otimes\mg\otimes \hs\,. 
\end{align}
These $\hs$-valued functions obviously contain the ordinary gauge fields associated with 
$s=0$ (or $\hat Y^{00}=\one$):
\begin{align}
  A_{\mu} = A_{\mu;\a}\l^\a \quad\in \End(\cH_{\cM})\otimes\mg\,.
\end{align}
The field strength in the presence of the higher-spin modes 
reads 
\begin{align}
    \cF^{\dot\mu\dot\nu}  = \cF^{\dot\mu\dot\nu;sm}_{\alpha}\lambda^{\alpha} \hat Y_{sm} \, ,
\end{align}
which again contains the ordinary field strength $\cF^{\dot\mu\dot\nu}_{\alpha}\lambda^{\alpha}$ for $s=0$.
Translating the frame (dotted) indices into spacetime (undotted) indices results in\footnote{We omit here a $\mathfrak{u}(1)$-valued contribution $d_\cK\theta^{-1}_{\mu\nu} \theta^{-1}_{\mu'\nu'} \gamma^{\mu\mu'} \gamma^{\nu\nu'}$ which is part of the geometric sector.}
\begin{align}
    \tr_{\cH_{\tJ}\otimes\cH_{\cK} 
    }\big(\cF^{\dot\mu\dot\nu}\cF_{\dot\mu\dot\nu}\big)
     = \tr_{\cH_{\tJ}\otimes\cH_{\cK}}\big(F_{\mu\nu} F_{\mu'\nu'}\big) \gamma^{\mu\mu'} \gamma^{\nu\nu'}
      = \rho^4 \tr_{\cH_{\tJ}\otimes\cH_{\cK}}\big(F_{\mu\nu} F_{\mu'\nu'}\big) G^{\mu\mu'} G^{\nu\nu'}\,,\qquad 
      \label{YM-term-contract-factor}
\end{align}
where $\rho$ is the dilaton. 
As usual, the trace $\tr_{\cH_{\tJ}\otimes\cH_{\cK}}$ factorizes as $\tr_{\cH_{\tJ}}\times\tr_{\cH_{\cK}}$. 
We shall normalize the generators $\lambda^{\alpha}$ in the Gell-Mann basis of $\mg \cong \msu(d_{\cK})$  in the standard way as 
\begin{align}
\label{lambda-normalization}
    \tr_{\cH_{\cK}}(\lambda^{\alpha}\lambda^{\beta})=2\,\delta^{\alpha\beta}\,.
\end{align}
On the other hand, we will use the normalization
\begin{align}
\label{normalization-Y-2}
    \tr_{\cH_{\tJ}}(\hat Y^{sm}\hat Y^{s'm'}) \approx \tJ\, \delta^{ss'}\delta^{m+m',0} \,
\end{align}
for the spherical harmonics $\hat Y^{sm} \in \hs \cong \msu(\tJ+1)$ of $S^2_{\tJ}$,
consistent with the normalization
$\hat Y^{00} = \one$ for the lowest-spin YM part.
These $\hat Y^{sm}$ satisfy the $\hs$ Lie algebra\footnote{We are using here a complexified basis of the real Lie algebra $\hs = \msu(\tJ+1)$. 
Of course the fluctuations $\cA^{\dot\mu}$ must be hermitian.}
\begin{align}
\label{structure-const-hs}
 [\hat Y^{sm},\hat Y^{s'm'}] = f^{sm;s'm'}_{kn}\hat Y^{kn} \ ,\qquad f^{sm,s'm',kn}=\frac{1}{\tJ}\tr_{\cH_{\tJ}}\Big([\hat Y^{sm},\hat Y^{s'm'}]\hat Y^{kn}\Big)\,.
\end{align}
The present normalization differs from the standard group theoretical normalization\footnote{Typically, one would like to normalize the spherical harmonics as
    $\tr_{\cH_{\tJ}}(\hat Y^{sm}\hat Y^{s'm'}) = \delta^{ss'}\delta^{m+m',0}$.}, to ensure that the lowest-spin $s=0$ mode takes the form
\begin{align}
    \cA^{\dot\mu} = 
   \cA^{\dot\mu}_{\a} \lambda^\a \one
\end{align}
so that these ``standard''  gauge fields couple appropriately via $[\lambda^\a,.]_{\mg}$
to the bi-fundamental fermions realized as string modes stretched between different branes. 

\paragraph{Effective couplings.} 
In the semi-classical late-time regime of the geometry,
the classical non-abelian (higher-spin)
Yang-Mills sector of the original action \eqref{SO(1,9)action} takes the following form
\begin{align}
    S_{\hs\text{-}\YM} &= -\frac 1{g^2} \int \mho_0\, \tr_{\cH_{\tJ}\otimes\cH_{\cK} }\big(\cF^{\dot\mu\dot\nu}\cF_{\dot\mu\dot\nu}\big)
    \sim -\frac{\tJ}{\ell_p^4 g^2} \int \limits_{\cM^{1,3}}\sqrt{G}\, \rho^2 \tr_{\cH_{\tJ}\otimes\cH_{\cK}}(F_{\mu\nu} F_{\mu'\nu'}) G^{\mu\mu'} G^{\nu\nu'} \ .
    \label{YM-action-bare}
\end{align}
Here we dropped an extra $U(1)$ term, which 
is part of the gravitational sector. Note that the explicit $\rho^2$ factor incorporates the cosmic expansion. In terms of above higher-spin modes, this classical action can be written as
\begin{align}\label{eq:classical-YM-HS}
    S_{\hs\text{-}\YM}&= -\sum_s\frac{\tJ}{\ell_p^4\,g^2}\int\limits_{\cM^{1,3}} \sqrt{G}\rho^{2}\tr_{\cH_{\cK}}\big(F_{\mu\nu;sm}F_{\mu'\nu'}{}^{sm}\big)G^{\mu\mu'}G^{\nu\nu'}\nn\\
    &=- 2\sum_s\int\limits_{\cM^{1,3}} \frac{\sqrt{G}}{g_{\YM}^2}  F_{\mu\nu;sm,\alpha}F_{\mu'\nu'}{}^{sm,\alpha}G^{\mu\mu'}G^{\nu\nu'}\,
\end{align}
where the Yang-Mills coupling is given by
\begin{align}
\label{YM-coupling}
   \frac 1{g_{\YM}^2} =  \frac{\tJ}{\ell_p^4\, g^2} \rho^2
   \propto
   \frac{\tJ}{\ell_p^4\, g^2} a^2(t)\,.
\end{align}
Here, $\ell_p^4\, g^2$ is dimensionless since the bare $g^2$ has dimension $[mass^4]$. This action may be interpreted as a $\hs$-extended or twisted 
Yang-Mills theory, where the ``original" gauge group $\tG:=SU((\tJ+1)\times d_\cK)$ of the model is transmuted to a \emph{finite-dimensional} higher-spin gauge subgroup $SU(\tJ+1)\subset \tG$.

Recalling that the dilaton $\rho$ 
grows with the cosmic expansion \eqref{rho-a(t)},
it follows that the YM coupling decreases with the cosmic evolution, i.e. $$g^2_{\YM}(t) \sim a(t)^{-2} \sim e^{-3\tau}\,,$$ in  agreement with the results of \cite{Steinacker:2023cuf}. 
This is an important observation, which  justifies the use of the 1-loop approximation in Section \ref{sec:4}. Note that $\frac 1{a(t)}\sim t^{-1}$ is slowly varying at late times, cf. \cite{Sperling:2019xar}.
The associated 't Hooft coupling is given by
\begin{align}
    \lambda_T=g_{\YM}\sqrt{\tJ} \propto \frac{\ell_p^2\,g}{a(t)}
\end{align}
where $\tJ$ drops out. It is important to note that this applies to the lowest-spin YM sector {\em and} the higher-spin sector. 
Therefore, the classical non-abelian couplings between higher-spin fields have the same strength as for the standard YM fields. 

\paragraph{Remarks on the $\hs$ couplings.} 
Typically, for a fuzzy 2-sphere $S^2_N$ encoded by some spherical harmonics $L^{sm}$ (for $s<N$) with a large cutoff $N$, the low spin sector are almost commuting as $f^{sm,s'm'}{}_{kn} \sim 1/\sqrt{N}$ cf. \eqref{structure-const-hs} while it is $f^{sm,s'm'}{}_{kn}\sim \sqrt{N}$ for some spin $s$ beyond the critical spin $s_*=\cO(\sqrt{N})$ where perturbation theory can not be trusted. However, this is not the case for the $S^2_{\tJ}$ in consideration. In particular, if we present 
\begin{align}
    \hat Y^{sm}=\hat Y_{\mu(s)}u^{\mu(s)}\,,\qquad u^{\mu}u_{\mu}=1\,,\qquad \{u^{\mu},u^{\nu}\}\sim \frac{1}{\tJ^2\cosh^2(\tau)}m^{\mu\nu}
\end{align}
for $u^{\mu}$ being the normalized fiber coordinates of $S^2_{\tJ}$, cf. Appendix \ref{app:A}. Then, it is clear that $\hat Y^{sm}$ are almost commutative even at large $s$. Thus, perturbation theory can be applied for generic $\mg\otimes\hs$-valued gauge fields in $\hs$-YM. 

There is another issue to be discussed. The present $\hs$ Yang-Mills theory is a rare example of a higher-spin theory which is approximately local and unitary with a truncated $\hs$ spectrum, and effectively Lorentz-invariant in $3+1$ dimensions.\footnote{Although mild apparent Lorentz-violation may occur for higher-derivative interactions as discussed in 
\cite{Steinacker:2023cuf}, Lorentz invariance is expected to be restored via gauge invariance.} Therefore, it is natural to ask how this theory surpasses standard no-go theorems \cite{Weinberg:1964ew,Coleman:1967ad} in the local physical regime, where we can use approximate plane-waves to study scattering amplitudes, see e.g. \cite{Steinacker:2023cuf,Steinacker:2023ntw}. 
We show in Section \ref{sec:massive} that the $\hs$-valued fields \eqref{eq:hs-valued-functions} in 
$\hs$-YM which carry the dof.  would-be-massive gauge fields encoded by the $\hat Y^{sm}$ with $2s+1$ dof.,\footnote{For a 
 careful analysis of the physical modes, see \cite{Steinacker:2019awe,Sperling:2019xar}.} can acquire mass through quantum effects; thus, leaving no tension with no-go theorems.

\section{The one-loop effective action}\label{sec:4}

We now study quantum effects from the one loop effective action of the IKKT matrix model with the setting of Section \ref{sec:2}. 



As in standard quantum field theory, the one-loop effective action in matrix model setting is obtained by integrating out (after gauge fixing) the fluctuating fields in the Gaussian approximation around the given background. This leads to \cite{Ishibashi:1996xs,Blaschke:2011qu,Steinacker:2023myp}
\begin{align}\label{eq:measure}
    Z_{\text{1-loop}}=\int dT d\Psi d\overline{\Psi} dcd\bar{c}\,e^{\im S_{\text{reg}}[T,\Psi,\overline{\Psi},c,\bar{c}]}=\exp\,\im (S[T]+ \Gamma_{\oneloop}[T])\,.
\end{align}
Here $S_{\text{reg}}$ is the IKKT action regulated by a suitable $\im\varepsilon$ term as in \cite{Steinacker:2023myp}.
The complete one-loop effective action of the IKKT model comprising the bosonic $(V)$, fermionic $(\Psi)$ and ghost contributions can be written as \cite{Blaschke:2011qu}:
\small
\begin{align}\label{Gamma-schwinger-susy}
\Gamma_{\oneloop}[T] 
&= \frac \im 2 \Tr \Bigg(\!\log\Big(\Box -\im\varepsilon -\Sigma^{(V)}_{\Ibold\boldJ}[\cF^{\Ibold\boldJ},-]\Big)
  - \frac 12 \log\Big(\Box -\im\varepsilon -\Sigma^{(\Psi)}_{\Ibold\boldJ}[\cF^{\Ibold\boldJ},-]\Big) 
 - 2 \log\Big(\Box-\im\varepsilon\Big) \! \Bigg)  \nn\\
&= -\frac \im 2 \int\limits_0^\infty \frac {d\a}{\a} 
   \Tr \Bigg( e^{-\im \a(\Box -\im\varepsilon - \Sigma^{(V)}_{\Ibold\boldJ}[\cF^{\Ibold\boldJ},-])}
  - \frac 12 e^{-\im\a(\Box -\im\varepsilon  -\Sigma^{(\Psi)}_{\Ibold\boldJ}[\cF^{\Ibold\boldJ},-]) }
 - 2 e^{-\im\a(\Box-\im\varepsilon)}  \Bigg)
\end{align}
\normalsize
using the identity\footnote{The oscillating integral is essential in Minkowski signature, since the real version would not be well-defined as $\Box_{1,3}$ is not positive.}
\begin{align}
\log\frac{Y+\im\varepsilon}{X+\im\varepsilon} \, &= 
\int_0^\infty\frac{d\a}{\a}\Big[e^{\im\a (X+\im\varepsilon)} - e^{\im\a (Y+\im\varepsilon)}\Big]\,.
\end{align}
Here, $\alpha$ is a Schwinger parameter, which captures the IR regime at $\alpha\rightarrow \infty$ and the UV physics at $\alpha\rightarrow 0$.
The trace $\Tr$ is over the space of all 
matrices or operators in the model, which will be convergent due to maximal supersymmetry.

Taking into account the structure $\cM^{1,3}\times S^2_{\tJ}\times  \cK$ of the background, we separate the space-time and internal contributions by writing $\Box=\Box_{1,3}+m_s^2+\Box_6$
, and splitting $\cF^{\Ibold\Jbold}$ according to \eqref{eq:F-10d}. Here
\begin{align}
   m_s^2=\frac{3(s+1)}{R^2}
\end{align}
are the IR masses of the $\hs$ modes  \cite{Sperling:2019xar}, which effectively tend to zero at late time. 
It is useful recalling that
\begin{align}
    \hs=\End(\cH_{\tJ})=\bigoplus_{s=0}^{\tJ}\hat Y^{sm}\,,\qquad -s\leq m\leq s\,,
\end{align}
contains 
harmonics $\hat Y^{sm}$ as 
irreducible representations of $\msu(2)$. 
To simplify the computation, we
assume that the non-abelian background associated with $\cK$ 
is (almost-) static. Then $\Box_{1,3}$ commutes\footnote{This is justified for the dominant part of $\Tr_{\Mat(\cH_{\cM})}$, where the kinetic contribution to $\Box_{1,3}$ dominates over the possible non-abelian components, which are assumed to be almost-constant.} with $\Box_{6}$
as well as $[\cF^{\Ibold\boldJ},-]$. 
We can then organize the full trace  as 
\begin{align}\label{eq:full-trace}
&\Tr_{\Mat(\cH)\otimes V}
\Big( e^{-\im\a(\Box -\im\varepsilon -\Sigma^{(V)}_{\Ibold\boldJ}[\cF^{\Ibold\boldJ},-])}\Big)\nn\\
&=\Tr_{\Mat(\cH_{\cM})\otimes \hs}\Big( e^{-\im\a(\Box_{1,3}+m_s^2-\im\varepsilon)} 
     \Tr_{\mg\otimes V}\big(
     e^{-\im\a(\Box_6  -\Sigma^{(V)}_{\Ibold\boldJ}[\cF^{\Ibold\boldJ},-])}\big)\!\Big)\,.
\end{align}
Assuming also that the mixed terms $\cF^{\dot\mu \jb}= 0$ vanishes (which is a reasonable assumption for slowly varying backgrounds), we 
can rewrite the bosonic contribution using
\begin{align}\label{tr-g-V-exp}
   \Tr_{\mg\otimes V}\Big(e^{-\im\alpha(\Box_6-\Sigma_{\Ibold\Jbold}^{(V)}[\cF^{\Ibold\Jbold,-}])}\Big)=\Tr_{\mg\otimes V_{(4)}}\Big( e^{\im\alpha(\Box_6- \Sigma^{(4)}_{\dot\mu\dot\nu}[\cF^{\dot\mu\dot\nu},-])}
     \Big)+\Tr_{\mg\otimes V_{(6)}}\Big(
     e^{-\im\a(\Box_6  -\Sigma^{(6)}_{\ib\jb}[\cF^{\ib\jb},-])}\Big)
\end{align}
where we have decomposed  $V = V_{(4)} \oplus V_{(6)}$
into the vector  representations of $SO(1,3)$ and $SO(6)$, respectively.

 Similarly, the 16-dimensional Weyl spinor representations of $SO(1,9)$ decomposes as 
\begin{align}
    \Psi_{(16)} = \Big(\Psi_{(2,-)} \otimes \Psi_{(4,-)}\Big)\oplus \Big(\Psi_{(2,+)} \otimes \Psi_{(4,+)}\Big)\,.
\end{align}
Analogous to the above,  $\Psi_{(2,\pm)}$ 
and $\Psi_{(4,\pm)}$
denote the chiral spinor representations of $SO(1,3)$ and $SO(6)$, respectively. 
It is useful to introduce the corresponding $SO(1,3)$ (${\rm ad}(\mg)$-valued) characters
\begin{align}\label{vector-Gamma-1-loop}
    \chi_{\cM^{1,3}}^{(4)}[\a] = \Tr_{V_{(4)}} \Big(e^{\im\,\a\Sigma^{(4)}_{\dot\mu\dot\nu}[\cF^{\dot\mu\dot\nu},-]} \Big)\,,\qquad
    \chi_{\cM^{1,3}}^{(2,\mp)}[\a] = \Tr_{\Psi_{(2,\mp)}} \Big(e^{\im\,\a\Sigma^{(2,\mp)}_{\dot\mu\dot\nu}[\cF^{\dot\mu\dot\nu},-]} \Big)\,.
\end{align}
The bosonic and fermionic contributions \eqref{tr-g-V-exp} can then be expressed as
\begin{subequations}
    \begin{align}
\Tr_{\mg\otimes V}\Big( e^{-\im\a(\Box_6 -\Sigma^{(V)}_{\Ibold\boldJ}[\cF^{\Ibold\boldJ},-])}\Big)
      &=  \Tr_{\mg\otimes V_{(6)}}\Big(e^{-\im\a(\Box_6  -\Sigma^{(6)}_{\ib\jb}[\cF^{\ib\jb},-])}\Big)
       +\Tr_{\mg}\Big(\chi_{\cM^{1,3}}^{(4)}[\a]e^{-\im\a\Box_6}\Big)\,,\\
 \Tr_{\mg\otimes \Psi}\Big( e^{-\im\a(\Box_6 -\Sigma^{(\Psi)}_{\Ibold\boldJ}[\cF^{\Ibold\boldJ},-])}\Big)
      &=+   \Tr_{\mg\otimes \Psi_{(4+)}}\Big(\chi_{\cM^{1,3}}^{(2+)}[\a]e^{-\im\a(\Box_6  -\Sigma^{(4+)}_{\ib\jb}[\cF^{\ib\jb},-])}\Big)  \nn\\
 & \quad  \, + \Tr_{\mg\otimes \Psi_{(4-)}}\Big(\chi_{\cM^{1,3}}^{(2-)}[\a]
     e^{-\im\a(\Box_6  -\Sigma^{(4-)}_{\ib\jb}[\cF^{\ib\jb},-])}\Big) \,.
\end{align}
\end{subequations}
For the specific background $\cM^{1,3}\times\cK$ in the last section, we can evaluate these traces in the basis of open-string modes on the branes, cf. \cite{Steinacker:2022kji,Steinacker:2023myp}.
It is convenient to first carry out the trace over $\End(\cH_\cM)$ 
by using \eqref{eq:Tr-End-M}. With this in mind,
\small
\begin{align}\label{eq:general-End-trace}
    \Tr\,\cO=\frac{1}{(2\pi)^4}\int_{\cM^{1,3}}\! d^4y\sqrt{G}\int\frac{d^4k}{\sqrt{G}}\Tr_{\mg\otimes \hs\otimes  V}(\cO)\,,
\end{align}
\normalsize 
where 
\begin{align}
    \cO=e^{-\im\a(\Box_{1,3}+m_s^2-\im\varepsilon)} f(y,\hat Y) 
\end{align}
is some $\mg\otimes \hs$-valued operator on $\cM^{1,3}$. 
We will mainly focus on the following integral:
\small
\begin{align}
\label{trace-spacetime-sector}
     &\Tr_{\Mat(\cH_{\cM})\otimes \hs}\Big( e^{-\im\a(\Box_{1,3}+m_s^2-\im\varepsilon)} f(y,\hat Y) \Big)= \frac{1}{(2\pi)^4} \int \limits_{\cM^{1,3}}\! \sqrt{G} 
      \int\frac{d^4 k}{\sqrt{G}}
    \Tr_{\hs}\Big( e^{-\im\a (k \cdot k+m_s^2)}f(y,\hat Y)\Big)
\end{align}
\normalsize
where
\begin{align}
  \Box_{1,3} = k\cdot k\,,\qquad k\cdot k=k_{\mu}k_{\nu}\gamma^{\mu\nu}=\rho^2k_{\mu}k_{\nu}G^{\mu\nu}\, 
\end{align}
arises from approximate plane waves on $\cM^{1,3}$ given by short string modes.
As a result, the 1-loop effective action can be written as follows 
\begin{align}
\label{eq:full-trace-HS-general}
\Gamma_{\oneloop}[T] 
&= -\frac 1{2(2\pi)^4} 
 \!\!\int\limits_{\cM^{1,3}} \!\! \sqrt{G} \rho^{-4}
 \int\limits_0^\infty \frac {d\a}{\a} \kappa[\a]\,
 \Tr_{\hs}\Bigg[  \Tr_{\mg\otimes V_{(6)}}\big(e^{-\im\a(\Box_6 +m_s^2  -\Sigma^{(6)}_{\ib\jb}[\cF^{\ib\jb},-])}\big)
     \nn \\
     &\quad  + \Tr_{\mg}\big(\chi_{\cM^{1,3}}^{(4)}[\a]
     e^{-\im\a(\Box_6+m_s^2)}\big)\nn\\
     &\quad -\frac 12 \sum_\pm \Tr_{\mg\otimes \Psi_{(4\pm)}}\big(\chi_{\cM^{1,3}}^{(2\pm)}[\a]e^{-\im\a(\Box_6 +m_s^2 -\Sigma^{(4\pm)}_{\ib\jb}[\cF^{\ib\jb},-])}\big) 
     - 2 \Tr_\mg \big( e^{-\im\a(\Box_6+m_s^2)}\big)\Bigg]\,
\end{align}
where $\kappa[\alpha]$, cf. \eqref{eq:kappa-a}, is obtained from carrying out the integral over $k$  using \eqref{trace-EndM-kappa}. 
This formula is exact under the stated assumptions;
note that the terms at leading order $\cO(\a^0)$ under the bracket cancel due to maximal SUSY.   Furthermore,
the $SO(1,3)$ characters can be evaluated in closed form if desired.

\section{Quantum \texorpdfstring{$\hs$}{hs}-YM at one loop}\label{sec:7}

We compute in this section the
effective coupling of the $\hs$-Yang-Mills gauge theory at one loop.  We also establish some requirements for obtaining a near-realistic gauge theory from the IKKT  model on the type of background under consideration.

As we will focus on the Yang-Mills term, which is quadratic in $\cF_{\dot\mu\dot\nu}$, we can expand  the $SO(1,3)$ characters \eqref{vector-Gamma-1-loop} 
using the following trace formulas for the $\mso(1,3)$ generators
\begin{subequations}\label{eq:various-trace-reps}
    \begin{align}
\Tr_{V_{(4)}}\Big(\Sigma^{V_{(4)}}_{\dot\mu\dot\nu}\Sigma^{V_{(4)}}_{\dot\rho\dot\sigma}\Big) &= 
2\big(\eta_{\dot\mu\dot\rho}\eta_{\dot\nu\dot\sigma} - \eta_{\dot\mu\dot\sigma}\eta_{\dot\nu\dot\rho}\big)\,,\\
\Tr_{V_{(6)}}\Big(\Sigma^{V_{(6})}_{\ib\jb}\Sigma^{V_{(6)}}_{\mb\nb}\Big)&=2\big(\delta_{\ib\mb}\delta_{\jb\nb}-\delta_{\ib\nb}\delta_{\jb\mb}\big)\,,\\
\Tr_{\Psi_{(2,\mp)}}\Big(\Sigma^{\Psi_{(2,\mp)}}_{\dot\mu\dot\nu}\Sigma^{\Psi_{(2,\mp)}}_{\dot\rho\dot\sigma} \Big)&= \frac 12(\eta_{\dot\mu\dot\rho}\eta_{\dot\nu\dot\sigma} - \eta_{\dot\mu\dot\sigma}\eta_{\dot\nu\dot\rho})\,,\\
\Tr_{\Psi_{(4,\pm)}}\Big(\Sigma^{\Psi_{(4,\pm)}}_{\ib\jb}
\Sigma^{\Psi_{(4,\pm)}}_{\kb\lb}\Big) &= \delta_{\ib\kb}\delta_{\jb\lb} - \delta_{\ib\lb}\delta_{\jb\kb}\,,
    \end{align}
\end{subequations}
where we have also included the $\mso(6)$ case for later use. This gives
\begin{subequations}
    \begin{align}
    \chi_{\cM}^{(4)}[\a]
    &=  4 -2\a^2[\cF_{\dot\mu\dot\nu},[\cF^{\dot\mu\dot\nu},]] + \cO(\a^4)\,,\\
    \chi_{\cM}^{(2,\mp)}[\a] &= 2-\frac{\a^2}{2}[\cF_{\dot\mu\dot\nu},[\cF^{\dot\mu\dot\nu},]] + \cO(\a^4)\,.
\end{align}
\end{subequations}
Then the contribution to the Yang-Mills action obtained from 
\begin{align}
   \Gamma_{\oneloop}^{\YM}[T]:= \Gamma_{\oneloop}[T]\Big|_{\cF_{\dot\mu\dot\nu}^2}
\end{align}
reads
\begin{align}\label{eq:G-YM-1}
\Gamma_{\oneloop}^{\hs-\YM}[T] 
&= -\frac 1{2(2\pi)^4} 
 \!\!\int\limits_{\cM^{1,3}} \!\! \sqrt{G} \rho^{-4}
 \int\limits_0^\infty d\a\,\a\, \kappa[\a] 
 \,\Tr_{\hs}\Big[ -2\Tr_{\mg}\Big([\cF_{\dot\mu\dot\nu},[\cF^{\dot\mu\dot\nu},-]]
     e^{-\im\a(\Box_6+m_s^2)}\Big)\nn\\
     &\quad +\frac 14 \sum_\pm \Tr_{\mg\otimes \Psi_{(4\pm)}}\Big([\cF_{\dot\mu\dot\nu},[\cF^{\dot\mu\dot\nu},-]]
     e^{-\im\a(\Box_6 +m_s^2 -\Sigma^{(4\pm)}_{\ib\jb}[\cF^{\ib\jb},-])}\Big) \Big]\,.
\end{align}
To proceed, we shall assume that the background $\cK$ is slowly varying so that $[\Box_6,\cF_{\ib\jb}] = 0$. Expanding 
\begin{align}
    \Tr_{\Psi_{(4\pm)}} e^{-\im\a(\Box_6+m_s^2  -\Sigma^{(4\pm)}_{\ib\jb}[\cF^{\ib\jb},-])}
    = e^{-\im\a(\Box_6+m_s^2)}
    \Big(4 - \a^2 [\cF_{\ib\jb},[\cF^{\ib\jb},-]] + \cO(\a^4) \Big)\,
\end{align}
in the UV regime and plugging this back to \eqref{eq:G-YM-1}, we observe that the leading contributions at order $\cO(\a^0)$ and $\cO(\a^2)$
cancel, thanks to maximal SUSY. The first non-trivial contribution starts at order $\alpha^4$, given by
\begin{align}\label{eq:Gamma-UV-prep}
\Gamma^{\hs-\YM}_{\oneloop}
&= +\frac 1{4(2\pi)^4} 
 \!\!\int\limits_{\cM^{1,3}} \!\!\sqrt{G} \rho^{-4}
 \int\limits_0^{\infty} d\a\,\a^3 \kappa[\a] 
 \Tr_{\mg\otimes \hs}\Big([\cF_{\dot\mu\dot\nu},[\cF^{\dot\mu\dot\nu},]]\,
  [\cF_{\ib\jb},[\cF^{\ib\jb},.]]e^{-\im\a(\Box_6+m_s^2)}\Big) 
\end{align}
dropping higher order contributions in $\cF$. 
Using 
\eqref{eq:kappa-a}, we learn that
\begin{align}
\label{int-alpha-1}
 \int_0^{\infty} d\alpha \,\alpha  e^{-\im\a (\Box_6+m_s^2)} 
  = -\frac{1}{(\Box_6+m_s^2)^2}\,.
\end{align}
Therefore,
\small
\begin{align*}
\Gamma^{\YM}_{\oneloop}
&= -\frac {1}{4(4\pi)^2} 
\!\!\int_{\cM^{1,3}} \!\! \sqrt{G} \rho^{-4}\Tr_{\mg\otimes \hs}\Bigg([\cF_{\dot\mu\dot\nu},[\cF^{\dot\mu\dot\nu},.]]\,
  [\cF_{\ib\jb},[\cF^{\ib\jb},.]]
  \frac{1}{(\Box_6+m_s^2)^2} 
  \Bigg) \,.
\end{align*}
\normalsize
The above can be simplified to 
\small
\begin{align}
\label{Gamma-1loop-YM-expand}
\Gamma^{\hs-\YM}_{\oneloop} 
&= -\frac {1}{4(4\pi)^2} 
\!\!\int\limits_{\cM^{1,3}} \!\! \sqrt{G} \rho^{-4}
\Tr_{\mg\otimes \hs}\Big(\frac{1}{\Box_6^2}[\cF_{\dot\mu\dot\nu},[\cF^{\dot\mu\dot\nu},.]]
  [\cF_{\ib\jb},[\cF^{\ib\jb},.]] 
  \Big) \,
\end{align}
\normalsize
under the assumption that $\Box_6\gg m_s^2$. 
Here, the trace over $\hs$-modes in the loop can be performed trivially. In particular, since $\hat Y$ are effectively commutative at late times, the trace $\Tr_{\hs}$ over the higher spin modes simply gives a factor $\tJ^2$
for large $\tJ$. Therefore
\begin{align}
    \Gamma^{\hs-\YM}_{\oneloop}
    \approx-\frac {\tJ^2}{4(4\pi)^2} \sum_{s,m}
\!\!\int\limits_{\cM^{1,3}} \!\! \sqrt{G} \rho^{-4}
\Tr_{\mg}\Bigg(\frac{1}{\Box_6^2}[\cF_{\dot\mu\dot\nu;sm},[\cF^{\dot\mu\dot\nu;sm},.]]
  [\cF_{\ib\jb},[\cF^{\ib\jb},.]] 
  \Bigg) \,.
\end{align}
Given that $\cF_{\dot\mu\dot\nu;sm}=\cF_{\dot\mu\dot\nu;sm\,\alpha}\l^{\a}$, we can use \eqref{YM-term-contract-factor} to write the above as
\begin{align}
\label{YM-1loop-UV-general}
\boxed{
    \Gamma^{\hs-\YM}_{\oneloop}=-\frac{\tJ^2}{4(4\pi)^2}\sum_{s,m}\int \limits_{\cM^{1,3}}\sqrt{G} \,G^{\mu\mu'}G^{\nu\nu'}  \langle F_{\mu\nu;sm},F_{\mu'\nu'}{}^{sm}\rangle_{\cK} 
    }\,
\end{align}
where
\begin{align}\label{eq:innter-product-UV}
     \langle \l,\mu\rangle_{\cK}
      := \Tr_{\mg}\Big(\frac{1}{\Box_6^2}[\l,[\mu,.]]\,  [\cF_{\ib\jb},[\cF^{\ib\jb},.]] 
  \Big)
\end{align}
is a modified inner product for any $\l,\mu \in \mg$. This inner product is non-negative due to the positivity of the operator $[\cF^{\ib\jb},[\cF_{\ib\jb},.]]$, and, in general, not $\mg$-invariant for non-trivial/reducible $\cK$. 

\paragraph{On separable branes.} Before working out some explicit examples, it is important to recall that if $\cK$ is non-trivial, there will be spontaneous breaking of $\mg$ symmetry (or $\mg$-invariance). In particular, consider
a configuration given by a set of separable branes $\cK_i$, where
\begin{align}\label{eq:cH-decompose}
    \cH_{\cK} = \bigoplus_{i=1}^n\cH_{\cK_i}  \,. 
\end{align}
\begin{figure}[ht!]
    \centering
    \includegraphics[scale=0.42]{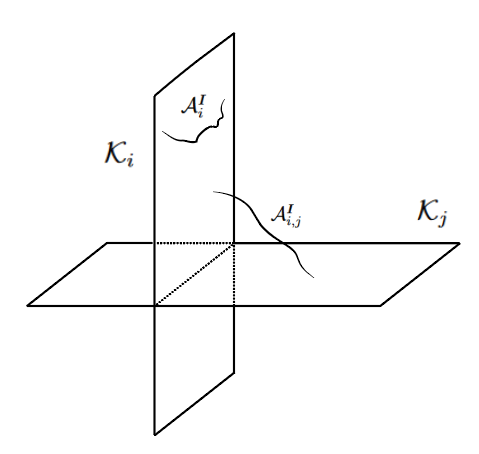}
    \caption{Open string modes $\cA^{\Ibold}_i$ start and end on the same brane $\cK_i$, while $\cA^{\Ibold}_{i,j}$ stretch between $\cK_i$ and $\cK_j$ branes.}
    \label{fig:openstrings-on-branes}
\end{figure}\\
In this case, the associated matrix algebra decomposes as
\begin{align}\label{eq:mg-decompose}
    \mg=\bigoplus_{i=1}^n\End(\cH_{\cK_i})\oplus\bigoplus_{i\neq j}\big(\cH_{\cK_i}\otimes \cH_{\cK_j}^*\big)\equiv \bigoplus_{i=1}^n\mg_i\oplus\bigoplus_{i\neq j}\big(\cH_{\cK_i}\otimes \cH_{\cK_j}^*\big)\,.
\end{align}
As a result, the potentials between different pair of branes can be computed by considering the appropriate Hilbert spaces. The decomposition \eqref{eq:mg-decompose} leads to two distinct types of 
fluctuations. On the one hand, when the fluctuations $\cA^{\Ibold}_i$ taking values in $\mg_i$, they can be interpreted as open strings starting and ending on the same brane $\cK_i$. On the other hand, $\cA^{\Ibold}_{i,j}$ taking values in $\cH_{\cK_i}\otimes \cH_{\cK_j}^*$ can be interpreted as open strings stretching between two different branes $\cK_i$ and $\cK_j$. In that case, the fluctuation $\cA^{\Ibold}_{i,j}$ takes values in the bi-fundamental representation $(d_{\cK_i},\bar{d}_{\cK_j})$ of $\cH_{\cK_i}\otimes \cH_{\cK_j}^*$. 

Using the decomposition \eqref{eq:mg-decompose}, the 
inner product $\langle .,.\rangle_\cK$ acquires various contributions from diagonal and off-diagonal blocks.
This can be evaluated quite concretely for 
$\cF^{\dot\mu\dot\nu}$ taking values in the unbroken gauge sector
$\mk\subset\mg$, i.e. the commutant of $\cK$, with generators $\l$. 
\\


$\symknight$ \underline{Scenario A: Point branes.} Consider the case where $\cK$ consists of $k$ coinciding point branes.
Then the unbroken gauge sector gets enhanced to
\begin{align}
   \mk =  \msu(k)\supset \mmu(1)_{1}\times\ldots \times \mmu(1)_{k} 
\end{align} 
leading to an unbroken $\msu(k)$ Yang-Mills gauge sector, which may be interpreted as color. 
\begin{figure}[ht!]
    \centering
    \includegraphics[scale=0.5]{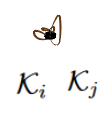}
    \caption{Coincident point branes, connected by open strings.}
\end{figure}\\
Furthermore, $\Box_6=0$ and $[\cF_{\ib\jb},[\cF^{\ib\jb},-]]=  0$,
so that $\langle .,.\rangle_\cK=0$. Therefore,
\begin{align}\label{eq:one-loop-point-brane}
    \Gamma^{\hs-\YM}_{\oneloop}(\text{point branes})=0\,.
\end{align}
This is to be expected, since there is no one-loop renormalization of the Yang-Mills coupling in the ordinary $\cN=4$ SYM, i.e. when $\cK$ is trivial.
Thus, the associated coupling of $\hs$-YM is given by the bare coupling \eqref{YM-coupling}. \\

$\symknight$ \underline{Scenario B: A stack of point branes and a bulk brane.} Now consider a stack of (coinciding) point branes denoted as
$\cK_k=\{p_1,\ldots,p_k\}$ and a large brane $\cK_j$. Generalizing the previous case, the unbroken gauge algebra is now $\mk = \mmu(k)$. 
We can then evaluate $\langle .,.\rangle_\cK$ 
as 
\begin{align}\label{trace-unbroken}
    \langle \l,\mu\rangle_{\cK}
    = \vartheta_{\mk} \tr_{\mg}(\l\mu) \, , \qquad \mbox{for} \ \ \l, \mu \in \mk\cong\msu(k)\,,
\end{align} 
which is proportional to the Killing form on $\mg$ up to some constant 
\begin{align}
\label{theta-mk-def}
     \vartheta_{\mk} = \frac 12 \Tr_{\mg}\Big(\frac{1}{\Box_6^2}[\l,[\l,.]]\,  [\cF_{\ib\jb},[\cF^{\ib\jb},.]]
  \Big) > 0\,,
\end{align}
We will use the normalization $\tr_\mg(\l\l) = 2$ and choose an adapted Gell-Mann basis $\l^\a\in \mg$ which include the generators in $\mk\cong\msu(k)$. Then,
\begin{align}
    \langle\l^\a,\l^\b\rangle_{\cK} = \vartheta_\mk \delta^{\a\b} \qquad \mbox{for} \ \ \l^\a \in \mk\,.
\end{align}
Due to string modes linking point branes with $\cK_j$,
there is a non-trivial contribution to  $\vartheta_\mk$.
To compute this, we take some $\l\in \mk$ as above, and 
evaluate the contribution from string modes
$\Upsilon = |i\rangle\langle z|$ connecting the point $i\in\cK_k$ with 
$\cK_j$ through a coherent state $|z\rangle$ on $\cK_j$. Since $\l$ act only on $|i\rangle$, we have
\begin{align}\label{eq:point-bulk-character}
   [\l,[\l,\Upsilon]] 
   = \l\l\Upsilon \ .
\end{align}
The integral over the string modes $\Upsilon$
in the geometric trace formula  \eqref{trace-coherent-End} amounts to an integral over points $z\in \cK_j$ and a sum over discrete points $i\in \cK_k$. The latter can be evaluated as
\begin{align}
    \sum_i \langle i| [\l,[\l,.]]|i\rangle
    =\sum_i\langle i| \l\l|i\rangle=\tr_{\mg}(\l\l) = 2\,.
\end{align}
\begin{figure}[ht!]
    \centering
    \includegraphics[scale=0.5]{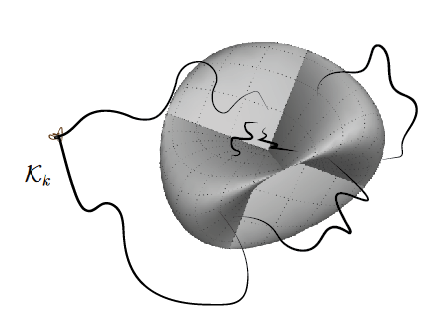}
    \caption{A stack of $k$ point branes interact with a bulk brane via open strings.}
\end{figure}
Now the trace over the large brane $\cH_{\cK_j}$ can be evaluated using \eqref{trace-coherent-End} if the geometry of $\cK_j$ is known explicitly, which gives
\begin{align}\label{theta-pointbrane-bulkbrane}
    \vartheta_\mk  \  &\sim \ 2\,
 \int_{\cK_j}\frac{\Omega_{z_j}}{(2\pi)^{\frac{|\cK_j|}{2}}} \frac{ \cF_{\ib\jb}(z_j) \cF^{\ib\jb}(z_j)}{(m^2_{\cK}z_j^2 + \Delta^2_{\cK})^2}\,.
\end{align}
Note that we integrate over \emph{dimensionless} variables, and all scales $(\Delta_{\cK},m_{\cK})$ are exhibited explicitly. 
It is also useful recalling that $z^2\sim \cO(1)$ and $m_{\cK}=d_{\cK}^{1/|\cK|}\Delta_{\cK}$ is the curvature radius of the branes, and  $\int\Omega_{\cK}=d_{\cK}$ from the Bohr-Sommerfeld quantization condition \eqref{Bohr-Sommerfeld}.

Putting this together, the unbroken $\msu(k)$-sector contributes an amount
\begin{align}
\label{suk-1loop}
     \Gamma^{\hs-\YM}_{\oneloop,\mk}=-\frac{\tJ^2\vartheta_{\mk}}{2(4\pi)^2}\int \limits_{\cM^{1,3}}\sqrt{G} \,G^{\mu\mu'}G^{\nu\nu'} F_{\mu\nu;\a},F_{\mu'\nu'}{}^{\a}
    \,
\end{align}
to the total one-loop effective action of the $\hs$-YM theory. We can compute $\vartheta_{\mk}$ quite explicitly 
using \eqref{theta-pointbrane-bulkbrane} and assuming $\cF_{\ib\jb}\approx const$ of order $\cO(\Delta^2_{\cK})$. In particular,
\begin{subequations}\label{vatheta-evaluated}
    \begin{align}
     |\cK_j|=2&:   &\vartheta_{\msu(k)}&\approx \frac{d_{\cK_j}}{2(1+d_{\cK_j})}\approx \frac{1}{2}\,,\\
     |\cK_j|=4&:    &\vartheta_{\msu(k)}&\approx \log(d_{\cK_j})>0\,,\\
     |\cK_j|=6&:    &\vartheta_{\msu(k)}&\approx \frac{d_{\cK}^{1/3}}{2} -\log(d_{\cK_j})>0\,
\end{align}
\end{subequations}
for large $d_{\cK_j}$. \\

$\symknight$ \underline{Scenario C: Large (extended) branes.} Now consider the case where $\cK$ consists of a number of large branes $\cK_i$. For simplicity, we restrict ourselves to the case of two large branes $\cK_1, \cK_2$ with associated Hilbert spaces $\cH_{\cK_i}$  cf. \eqref{eq:cH-decompose}. 
This sector receives significant contributions 
due to the long/heavy open string modes $\Upsilon = |z_1\rangle\langle z_2|\in \cH_{\cK_1}\otimes \cH_{\cK_2}^*$ linking $\cK_1$ with $\cK_2$ for $|z_i\rangle \in \cH_{\cK_i}$ being coherent states on each brane.

The unbroken gauge group in this case is generated by $\mk = \mmu(1)_1 \times \mmu(1)_2$, 
where the subscripts indicate the associated branes. Note that the inner product $\langle \l,\l\rangle_\cK$ cf. 
\eqref{eq:innter-product-UV} for $\l=\diag(\one_1,-\one_2) \in \mk $ can be evaluated as 
\begin{align}
    [\l,|z_1\rangle\langle z_2|] =  2|z_1\rangle\langle z_2|\,.
\end{align}
Furthermore, unlike scenario B), the normalization for the $\l$ generators in this case is given by
\begin{align}
    \tr_\mg(\l \l) = d_{\cK_1} + d_{\cK_2}\,.
\end{align}
Analogously to \eqref{theta-mk-def}, 
we write
\begin{align}
    \langle \l,\l\rangle_\cK = \frac 12\vartheta_\mk^{\cK_1-\cK_2} \tr_\mg(\l\l)\,.
\end{align}
Then the geometric trace formula yields
\begin{align}\label{eq:geo-trace}
 \vartheta_\mk^{\cK_1-\cK_2} \sim \frac{8}{d_{\cK_1} + d_{\cK_2}}
     \int\limits_{\cK_1\times \cK_2}\frac{\Omega_{z_1} \Omega_{z_2}}{(2\pi)^{\frac{|\cK_1|+|\cK_2|}{2}}} \frac{ (\cF_{\ib\jb}(z_1) - \cF_{\ib\jb}(z_2))  (\cF^{\ib\jb}(z_1) - \cF^{\ib\jb}(z_2))}{\big(m^2_{\cK}(z_1-z_2)^2 + \Delta^2_{\cK}\big)^2}\,.
\end{align}
\normalsize
Since $\vartheta_\mk^{\cK_i-\cK_j}$ will be large as $\cK_i$ and $\cK_j$ are large, the effective couplings \eqref{renorm-YMcoupling} associated with this sector will typically be suppressed in the low-energy regime.  \\

$\symknight$ \underline{Scenario D: Self-intersecting branes.} The formula \eqref{eq:geo-trace} can essentially be used also for self-intersecting branes such as squashed $\P^2_*$ cf. \cite{Steinacker:2015dra},
where the intersecting sheets of these type of  branes play (approximately) the role of different $\cK_i$.
There are also contributions from 
 (almost-) zero modes arising as string modes  $\Upsilon = |z_i\rangle\langle z_j|$. Here, $|z_j\rangle$ are the associated 
coherent states located at the intersection of the different sheets of the brane. 
They may be significant because 
$\cF_{\ib\jb}(z_i) - \cF_{\ib\jb}(z_j)$
does not necessarily vanish at the intersection locus. 
Such considerations may be relevant to construct analogs of $W$ or $Z$ bosons, since chiral fermions taking values in the bi-fundamental representations of $\cH_{\cK_i}\otimes \cH_{\cK_j}^*$ arise naturally at the intersections. 
\begin{figure}[ht!]
    \centering
    \includegraphics[scale=0.5]{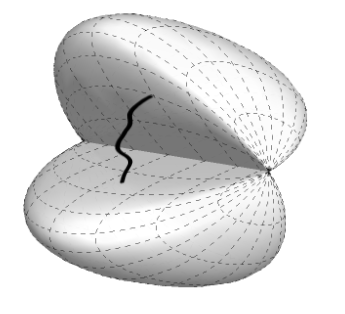}
    \caption{An open string stretching between two sheets of a self-intersecting brane.}
    \label{fig:self-intersecting-sphere}
\end{figure}



\paragraph{Effective coupling.} Combining the  one-loop contribution (for various cases above)  with the classical action, we obtain the effective\footnote{Recall that since the underlying theory is UV finite, no renormalization procedure 
is required.} action of the quantum $\hs$-Yang-Mills theory as 
\small
\begin{align}
\label{effective-action-YM-wetterich}
    S^{\rm eff}_{\hs-\YM}&\approx\int \limits_{\cM^{1,3}} \Bigg(\frac{2\tJ}{g^2}\rho^2 +\frac{\vartheta_{\cK}\tJ^2}{4(4\pi)^2}
    \Bigg)\cL_{\hs-\YM}^0\,
\end{align}
\normalsize
where
\begin{align}
    \cL^0_{\hs-\YM}:= -\sum_{s,m}\sqrt{G}F_{\mu\nu;sm\,\alpha}\tensor{F}{_{\mu'}_{\nu'}_{sm}^\alpha}G^{\mu\mu'}G^{\nu\nu'}\,.
\end{align}
Then 
the effective coupling constant $g^R_{\YM}$ for the quantum Yang-Mills theory induced by the IKKT matrix model reads
\begin{align}
    \frac{1}{(g^R_{\YM})^2}=&\frac{2\tJ }{g^2}\rho^2+\frac{\vartheta_{\cK}\tJ^2}{4(4\pi)^2}\,
\end{align}
or equivalently
\begin{align}\label{renorm-YMcoupling}
    g^R_{\YM}\approx \frac{1}{\sqrt{\frac{2\tJ }{g^2}\rho^2+\frac{\vartheta_{\cK}\tJ^2}{4(4\pi)^2}}}\, .
\end{align}
\normalsize

\begin{figure}[ht!]
    \centering
    \includegraphics[scale=0.52]{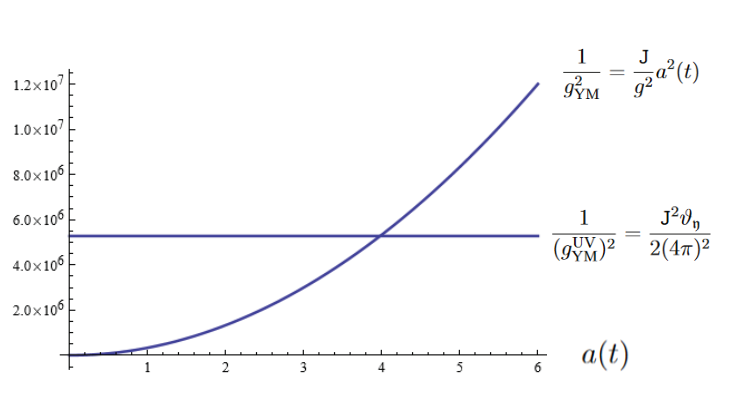}
    \caption{
For a sufficiently early stage of the cosmic evolution, the one-loop effective action dominates the classical action, leading to an approximate constant coupling constant. After some characteristic time, the interaction strength decreases in time, as $a(t) \propto \rho$ starts to dominate.}
    \label{couplings}
\end{figure}



\paragraph{Discussion.} 

To understand the significance of \eqref{renorm-YMcoupling}, 
we recall that $\rho \propto a(t)$ cf. \eqref{rho-a(t)}  increases with the cosmic expansion.
Therefore, for sufficiently early times, the renormalized coupling $g^R_{\YM}$ is dominated by the 1-loop contribution and is
approximately constant. More specifically, it is
much smaller 
than the bare coupling $g_{\YM}$.
At some characteristic time in the cosmic evolution, both the bare and 1-loop contributions become equal. Afterward, the  bare action starts to dominate, so that the effective coupling decreases in time.\footnote{Analogous effects can be expected for the gravitational sector of the matrix model, cf. \cite{Steinacker:2023myp}.} 

\section{Massive (higher-spin) YM gauge fields at one loop}\label{sec:massive}

Recall that, on a $\cM^{1,3}$ background, all fluctuations in the IKKT model are part of some (truncated) tower of higher-spin modes, which arise from a ``hidden" $S^2_{\tJ}$ factor of $\cM^{1,3}$. 
However, massless higher-spin YM gauge fields would be phenomenologically unacceptable, since they would interact significantly and be
generated in all sorts of scattering processes.
Therefore, there should be some mechanism for the $\hs$ fields in the unbroken gauge sector to acquire mass, so that we are \emph{not} in conflict with no-go theorems \cite{Weinberg:1964ew,Coleman:1967ad}. We shall exhibit such a mechanism  in this section, via quantum effects.


As a first hint towards such a mechanism, we observe that the higher-spin fields in $\hs$-YM theory typically carry all dof. of massive  fields \cite{Steinacker:2019awe,Sperling:2019xar}. Hence, it is natural to expect them to acquire mass in some way. We will show that all (higher-spin) Yang-Mills gauge fields indeed acquire mass at one loop in the presence of $\cK$. These quantum effects should \emph{not} be confused with the standard Higgs mechanism, since it arises only at quantum level. Note that these ``quantum'' mass arises 
even for the unbroken sector of the gauge group, and comprises a priori contributions with both signs.


While the present mechanism allows us to avoid no-go theorems, we observe that the mass of the 
 lowest-spin YM gauge bosons 
 is {\em larger} than their $\hs$ siblings.
This is somewhat disappointing, 
since we can then no longer use the present setting to reproduce the standard massless Yang-Mills gauge theory
. It remains to be seen whether this issue can be circumvented by other mechanisms\footnote{One possibility is to consider the minimal quantum space with $\tJ=0$, which will be elaborated elsewhere.}.

To obtain the induced mass term, consider a $\mg\otimes\hs$-valued field with internal spin-$s$ represented as
\begin{align}
\label{color-gauge-hs-ansatz}
    \cA_{\dot\mu}^{(s)} = \cA_{\dot\mu;\a s}\l^\a \hat Y^{s}\,,
\end{align}
where the $m$ index in $\hat Y^{sm}$ is dropped for better readability. We observe that there are two contributions to the mass term:
\begin{itemize}
    \item The first contribution arises from the kinetic term $\Box_{1,3}$ in the one-loop effective action cf. \eqref{trace-spacetime-sector} expanded to second order in fluctuations. This contribution leads to almost the same mass for the lowest-spin and higher-spin fields. Since the mass induced by this contribution is typically very large, certain type of brane configurations are more favorable than the others. 
    \item The second arises from the 
    $[\cF_{\dot\mu\dot\nu},[\cF^{\dot\mu\dot\nu},.]]\vartheta_{\mk}$ contribution for \emph{pure} $\hs$-valued flux. 
    This contribution sets the $\hs$-valued fields apart from their lowest-spin counterparts, but in the ''wrong`` direction. In particular, the mass induced by $\hs$-valued flux is negative, which can only be stabilized by the contributions coming from the kinetic term.
\end{itemize}

\subsection{Kinetic contribution}
\label{sec:mass-kinetic-UV}

We start with the first contribution to the mass term, obtained from the quadratic term in $\cA$ from an expansion of  $\Gamma_{\oneloop}[\bar T+\cA]$ around $\cA=\text{const}$. 
In this background, the space-time matrix d'Alembertian operator gets shifted to
\begin{align}
    \Box_{1,3} = \Box_{1,3}^0 + \Box_\cA, \qquad  \Box_\cA = [\cA^{\dot\mu},[\cA_{\dot\mu},.]]
\end{align}
where $\Box^0_{1,3}$ is the bare (3+1)-dimensional d'Alembert operator. Note that $\Box_\cA$
acts non-trivially on both $\hs$ and $\mg$ sectors. Furthermore, the $\hat Y$ can be considered as commutative functions  at late times (cf. Appendix \ref{app:A}). Then,
\begin{align}
    \Box_\cA = [\cA^{\dot\mu},[\cA_{\dot\mu},.]]
     \approx \cA^{\dot\mu}{}_{\a s}\cA_{\dot\mu;\b s'}\hat Y^{s} \hat Y^{s'} [\l^\a ,[\l^\b,.]] \ .
\end{align}
The trace $\Tr_{\hs}$ in \eqref{trace-spacetime-sector} including $\Box_\cA$ can be evaluated 
by noting that
 \begin{align}
     \Tr_{\hs}\Big(\hat Y^{sm}\hat Y^{s'm'} [\lambda_\a ,[\lambda_\b ,-]]\cO\Big)  
     &\approx   \tJ^2\delta^{ss'}\delta^{m+m',0}[\lambda_\a ,[\lambda_\b ,-]]\cO\,,\label{hs-trace-comm}
\end{align}
where the $\hat Y$ are normalized as in \eqref{normalization-Y-2}. 
Here, $\cO=\cO(\Box_{1,3},\Box_6,[\cF_{\ib\jb},[\cF^{\ib\jb},-]])$ is some operator assumed to act trivially on the $\hs$-modes to a good approximation. This amounts to the replacement
\begin{align}
    \Tr_{\hs}\Box_\cA  \ \to \  \tJ^2 \cA_{{\dot\mu};\a s} \tensor{\cA}{^{\dot\mu}_{\b s}}  [\l^\a,[\l^\b,.]]\, .
\end{align}
Combining the above results, we obtain from \eqref{trace-spacetime-sector} 
\begin{align}\label{eq:hs-trace-massive}
 &  \Tr_{\Mat(\cH_{\cM})\otimes\hs\otimes\mg}\Big( e^{-\im\a(\Box_{1,3}^0 + \Box_\cA -\im\varepsilon)} f(y) \Big)\nn\\
      &\approx -\im \tJ^2
      \frac{\kappa[\a]}{(2\pi)^4} \int_{\cM^{1,3}}\frac{\sqrt{G}}{\rho^{4}} f(y) \Tr_\mg\big(1 -\im \a\cA_{{\dot\mu};\a s} \tensor{\cA}{^{\dot\mu}_{\b s}}
      [\l^\a,[\l^\b,.]]\big) + \cO(\cA^4) \,
\end{align}
\normalsize
using \eqref{trace-EndM-kappa}.
To cast the mass terms in more standard form, we rewrite $\cA^{\dot\mu} = E^{\dot\mu \mu} A_\mu$ using the frame in terms of a 
one-form valued field $A_\mu$ as in \eqref{A-frame}.
This gives
\begin{align}
\cA^{\dot\mu}_{\a s}\cA_{\dot\mu;\b}{}^{s}
 = \gamma^{\mu\nu}
 A_{\mu;\a s}A_{\mu';\b}{}^{s}  
 = \rho^{2} 
G^{\mu\mu'}
 A_{\mu; \a s} A_{\mu';\b}{}^{s} ,\qquad
\end{align}
recalling \eqref{eq:eff-G}.
Assuming that $\cA$ takes values in some unbroken gauge algebra $\mk\cong \msu(k)$ cf. Section \ref{sec:7}, then
\begin{align}
    \Tr_{\mg}([\l^\a,[\l^\b,.]] \cO) = 
    \frac{\delta^{\a\b}}{\dim\mk}\Tr_{\mg}( C^2[\mk] \cO)
\end{align}
using $\mk$- invariance, 
where 
\begin{align}
    C^2[\mk] := [\l_\a,[\l^\a,.]]
\end{align}
 is the Casimir of $\mk$ acting in the adjoint on $\mg$.
 Then
 \eqref{eq:hs-trace-massive} leads to the 
 following contribution to the effective action
\begin{align}
\Gamma_{\oneloop}[\bar T+\cA] 
&= 
 \Gamma_{\oneloop}[\bar T] 
 - \frac {1}{(2\pi)^4} 
 \!\!\int\limits_{\cM^{1,3}} \!\! \sqrt{G} \, (m_{\rm 1-loop}^{\kin})^2  G^{\mu\mu'}A_{\mu;\a s}A_{\mu'}{}^{\a s}+\cO(A^4) \ 
\end{align}
with 
the induced mass for the $\hs$-valued $\msu(k)$ gauge fields is given by
\begin{align}\label{eq:m-kinetic-1-loop}
\boxed{
 (m_{\rm 1-loop}^{\kin})^2  
 \approx -\im \, \frac { \tJ^2}{\rho^{2}\dim\mk }
 \int\limits_0^\infty d\a\, \kappa[\a]\,
  \Tr_{\mg}\big( C^2[\mk]\,
 e^{-\im\a\Box_6}
 \sX^{(6)}\big) \ \stackrel{!}{>} \ 0 \,
 }
\end{align} 
assuming $[\Box_6,\cF_{\ib\jb}] = 0$ as usual.
Note that $m^2_{\oneloop}$ is real  since the integral over $\a$ produces another factor of $\im$, and it has the correct dimension since $\a$ has dimension $[mass^{-2}]$.
Here,
\begin{align}\label{eq:character-K-scalar-UV}
     \sX^{(6)}[\cF] &:= 2+
\Tr_{V_{(6)}}\big(e^{\im\a\,\Sigma^{(6)}_{\ib\jb}[\cF^{\ib\jb},-]}\big)
 - \sum_\pm \Tr_{\Psi_{(4\pm)}}\big(e^{\im\a\,\Sigma^{(4\pm)}_{\ib\jb}[\cF^{\ib\jb},-]}\big)\nn\\
 &= +\alpha^4\Big(4 [\cF^{\ib\jb},[\cF_{\jb\kb},[\cF^{\kb\mb},[\cF_{\mb\ib},-]]]]-[\cF^{\ib\jb},[\cF_{\ib\jb},[\cF^{\mb\nb},[\cF_{\mb\nb},-]]]]\Big) +\cO(\alpha^6)\,,
\end{align}
which will also govern the scalar potential. Integrating over $\a$ and keeping only the $\a^4$ term, we obtain\footnote{Note that this would vanish for trivial $\cF^{\Ibold\Jbold}$, i.e. without fuzzy extra dimensions $\cK$. }
\begin{align}\label{eq:negative-mass}
   (m_{\rm 1-loop}^{\kin})^2 \approx \frac{2\tJ^2\pi^2}{\rho^2\dim\mk}\Tr_{\mg}\Big(C^2[\mk]\frac{4\Tr(\delta \cF^4)-(\Tr(\delta \cF^2))^2}{\Box_6^3}\Big) \ \stackrel{!}{>} \ 0 \ .
\end{align}
where $\delta\cF = [\cF,.]$. The trace over $\mg$ can now be evaluated using the geometric trace formula. 
To be specific, we will consider the following two scenarios:
\begin{itemize}
\item[A)] a stack of $k$ point branes and a bulk brane as in Sections \ref{sec:7}
and \ref{sec:5},  and 

\item[B)]
a stack of $k$ coincident large bulk branes.
\end{itemize}
Both cases lead to an unbroken gauge algebra $\mk = \msu(k)$  
identified as the conventional ``color",\footnote{At the classical level, this $\msu(k)$ gauge theory would comprise a tower of $\hs$-valued massless higher-spin color gauge fields, which is phenomenologically unacceptable. } 
but the generators $\l$ are small as functions in A), but large as functions in B). This 
leads to some qualitatively different behavior.\\

\underline{Scenario A.} In this case, the off-diagonal string modes $|i\rangle\langle y|$ linking the point branes and $\cK$ will 
lead to
\begin{align}
\label{color-mass-bulk}
    (m_{\rm 1-loop}^{\kin})^2  
    &\approx \frac{2\tJ^2\pi^2}{\rho^{2}}
 \int\limits_{\cK}\! \frac{\Omega_y}{(2\pi)^{|\cK|/2}}\,\frac{4(\Tr\cF^4)-(\Tr(\cF^2))^2}{(m^2_{\cK}y^2+\Delta_{\cK}^2)^3}\, = \cO(\Delta_\cK^2)
\end{align}
recalling that $\cF = \cO(\Delta_\cK^2)$
where $\Delta^2_{\cK}$ is the scale of non-commutativity on $\cK$.

The explicit value for this induced mass depends significantly on the structure of $\cK$.
We can  estimate its value depending on the dimension of $\cK$ 
as follows 
\begin{subequations}
\label{gauge-masses-kinetic-dim}
    \begin{align}
        |\cK|=2 &:  &(m_{\rm 1-loop}^{\kin})^2 & \propto 
        \frac{\tJ^2\,
        }{\rho^2} 
        \Delta_\cK^2  \,,\\ 
        |\cK|=4&: &(m_{\rm 1-loop}^{\kin})^2 &\propto  \frac{\tJ^2\,
        }{\rho^2} 
        \Delta_\cK^2, \\
        |\cK|=6&:  &(m_{\rm 1-loop}^{\kin})^2 & \propto \frac{\tJ^2\,
        }{\rho^2} 
        \log d_\cK \Delta_\cK^2\,,
    \end{align}
\end{subequations}
assuming large $d_{\cK}$, and recalling $m_{\cK}=d_{\cK}^{1/|\cK|}\Delta_{\cK}$. Here $\Delta_\cK$ is a characteristic KK scale on $\cK$ \eqref{scales-K},
and therefore assumed to be in the UV regime from the space-time point of view. 
We observe that for $\dim\cK \leq 4$, the integral 
in \eqref{eq:negative-mass} is dominated by local contributions, and therefore almost vanishes for (anti-) self-dual $\cF_{\ib\jb}$,
since the contribution to the mass from very long strings on $\cK$ is small.
This scenario is realized for self-intersecting 4-dimensional branes such as $\P^2_*$ \cite{Steinacker:2015dra}.
Then the 1-loop contributions to the $\mk$-valued gauge bosons is small, both for conventional and $\hs$-valued gauge fields.\\


\underline{Scenario B.} Consider $k$ coincident large branes $\cK$. The unbroken algebra $\mk$ is generated by 
$\l_{ij} = \one_i \otimes \one_j$ where $\one_i$ is the projector on brane $\cK_i$.
We observe that all strings starting and ending on the same brane commute with $\l$ and therefore drop out. Hence, 
the non-trivial modes contributing to $\Tr_\mg$ consist of string modes linking different branes $i$th with $k$th,
which satisfy $\l_{ij}|j\rangle\langle k| \sim |i\rangle\langle k|$ etc. Then $C^2[\mk]=[\l_{\a},[\l^\a,.]]$ provides a $k$ on such 
string modes. As such, we obtain 
\begin{align}
   (m_{\rm 1-loop}^{\kin})^2 &\approx \frac{2\tJ^2\pi^2 k}{\rho^2\dim\mk}\Tr_{\mg}\Big(\frac{4\Tr(\delta \cF^4)-(\Tr(\delta \cF^2))^2}{\Box_6^3}\Big) \ \nn\\
 &\approx \frac{2\tJ^2\pi^2 k}{\rho^2}
  \int\limits_{\cK\times\cK}\! \frac{\Omega_x\Omega_y}{(2\pi)^{|\cK|}}\,\frac{4\Tr(\cF(x)-\cF(y))^4-(\Tr(\cF(x)-\cF(y))^2)^2}{(m^2_{\cK}(x-y)^2+\Delta_{\cK}^2)^3} \stackrel{!}{>} \ 0\,
\end{align}
where the integral is over a pair of 
the branes under consideration.
The scale of this mass can be estimated as in \eqref{gauge-masses-kinetic-dim}, but there will be an extra overall factor $d_\cK = \dim\cH_\cK$ due to the double integral.

We recall that 
the above calculations apply to all $\hs$ modes, 
since the $\hat Y^{sm}$ can be considered as classical functions. There is no significant distinction between the 
$s=0$-sector and the $s\geq 1$-sector from the kinetic contribution. 

\subsection{Flux contribution}
\label{sec:flux-mass}

Now consider the contribution of the flux term $\delta \cF_{\dot\mu\dot\nu}^2\delta\cF_{\ib\jb}^2$ 
to the mass. 
We can assume that 
$\cA^\nu = \cA^\nu_{s\a} \l^\a \hat Y^{s}$
where $\cA^\nu_{s\a}$ is constant, and $\hat Y^{s}$ are polynomials of order $s$ in $u^i$. For $s=0$, it follows immediately that $\cA$ drops out from $\cF_{\dot\mu\dot\nu}$, so that the induced mass vanishes.
This means the flux contribution distinguishes the $\hs$ sector from the conventional $s=0$ sector. However, this contribution is shown to go in the wrong direction as alluded to in the above.
For simplicity, consider $s=1$ case where 
$\cA^\nu = \cA^\nu_{i\a} \l^\a u^i$. Then,
\begin{align}\label{eq:flux-mass}
\cF_{\dot\mu\dot\nu} &= 
\{\ttb_{\dot\mu},\ttb_{\dot\nu}\}+ \{\ttb_{\dot\mu},\cA_{\dot\nu}^{(1)}\}+ \{\ttb_{\dot\nu},\cA_{\dot\mu}^{(1)}\}+\im [\cA_{\dot\mu}^{(1)},\cA_{\dot\nu}^{(1)}] \nn\\
&=
\{\ttb_{\dot\mu},\ttb_{\dot\nu}\}+ \cA_{\dot\nu i\a}\{\ttb_{\dot\mu},
 u_i\} \l^\a  
 + \cA_{\dot\mu i\a}\{\ttb_{\dot\nu},u_i\}\l^\a 
 +\im \cA_{\dot\mu i\a}
 \cA_{\dot\nu j\b}[\l^\a u_i,\l^\b u_j] \nn\\
  &\approx \frac{1}{R^2} m_{\dot\mu\dot\nu} + 
  \frac{\ell_p}{R^2\cosh(\tau)}\big(\cA_{\dot\nu i\a} m_{\dot\mu i}  
 + \cA_{\dot\mu i\a} m_{\dot\nu i}\big) \l^\a 
 +\im \cA_{\dot\mu i\a}
 \cA_{\dot\nu j\b} f^{\a\b}_\g u_iu_j  \l^\g 
 \,.
\end{align}
where for convenience, we recall that\footnote{For further information, see Appendix \ref{sec:S2J-structure}.}
\begin{align}
    y^2&=-R^2\cosh^2(\tau),\qquad \ttb^2=\ell_p^{-2}\cosh^2(\tau),\qquad  u^{\mu}=\frac{\ell_p}{\cosh(\tau)}\ttb^{\mu}\,.
\end{align}
Here, the middle terms in \eqref{eq:flux-mass} lead to the desired mass term in $[\cF_{\dot\mu\dot\nu},[\cF^{\dot\mu\dot\nu},.]]$, while
the last term will drop out. Therefore, if we are interested in estimating the magnitude of the mass term, then effectively
\begin{align}
    \cF_{\dot\mu\dot\nu}  &\approx  
  \frac{\ell_p}{R^2\cosh^2(\tau)}\big(\cA_{\dot\nu i\a} m_{\dot\mu i}  
 + \cA_{\dot\mu i\a} m_{\dot\nu i}\big) \l^\a \,. 
\end{align}
As a result,
\begin{align}
    [\cF_{\dot\mu\dot\nu},[\cF^{\dot\mu\dot\nu},.]]
    \approx \frac{\ell_p^2\rho^2}{R^2\cosh^4(\tau)}A_{\mu i\a}A^{\mu}{}_{j\b}\Big(m_{\dot\nu}{}^im^{\dot\nu j}[\l^a,[\l^{\b},.]]+\l^{\a}\l^{\b}[m_{\dot\nu}{}^i,[m^{\dot\nu j},.]]\Big)\,,
\end{align}   
where $[m,m]$ contributions are subleading
. Note that the above two terms have opposite signs. In particular,
\begin{align}
    m_{\dot\nu}{}^im^{\dot\nu j}[\l^a,[\l_{\a},.]]<0\,,\qquad \l^{\a}\l^{\b}[m_{\dot\nu i},[m^{\dot\nu i},.]]>0\,,
\end{align}
where $[m_{\dot\nu j},[m^{\dot\nu j},.]]>0$ contains the Casimir of $SO(3)$ with standard normalization. Using various relations in Appendix \ref{app:A}, we can simplify the above to
\begin{align}
    [\cF_{\dot\mu\dot\nu},[\cF^{\dot\mu\dot\nu},.]]&\approx A_{\mu i\a}A^{\mu}{}_{j\b}\Big(-\frac{\ell_p^2\rho^2}{\cosh^2(\tau)}\ttb^i\ttb^j[\l^\a,[\l^{\b},.]]+\frac{\ell_p^2\rho^2}{R^2\cosh^4(\tau)}\l^\a\l^\b[m_{\dot\nu}{}^i,[m^{\dot\nu j},.]]\Big)  \nn\\
    &\approx\rho^2 A_{\mu i\a}A^{\mu}{}_{j\b}\Big(-u^i u^j[\l^\a,[\l^{\b},.]]+\frac{4}{\tJ^2\cosh^4(\tau)}\l^\a\l^\b[m_{\dot\nu}{}^i,[m^{\dot\nu j},.]]\Big)
\end{align}  
where we have restricted ourselves to the reference point $\tp$ where $y^i=y^{\mu}\big|_{\tp}=(y^0,0,0,0)$. 
Note that 
\begin{align}
    \Tr_\hs [m_{\dot\mu}{}^i,[m^{\dot\mu j},.]] \sim \tJ^4 \delta^{ij}\,,\qquad  \Tr_\hs (u^i u^j) \sim  \tJ^2\delta^{ij}\,.
\end{align}
Then, from \eqref{Gamma-1loop-YM-expand}, we obtain the following contribution to the effective action
\small
\begin{align}
    \eqref{Gamma-1loop-YM-expand}\Big|_{A^2}\approx-\frac{\tJ^2}{4(4\pi)^2}\int\limits_{\cM^{1,3}}\frac{\sqrt{G}}{\rho^2}A_{\mu i \a}A^{\mu i}{}_{\b}\Tr_{\mg}\Bigg(\frac{1}{\Box_6^2}\Big(\frac{4\,\l^\a\l^\b}{\cosh^4(\tau)} 
    -[\l^{\a},[\l^\b,.]]\Big)[\cF_{\ib\jb},[\cF^{\ib\jb},.]]\Bigg)\,.
\end{align}
\normalsize
As always, the trace $\Tr_\mg$ can be evaluated
using 
\eqref{eq:general-End-trace}. If $\l^\a$ are generators in the unbroken gauge group with low rank, then both $\l^\a \l^\b$ and $[\l^\a,[\l^\b,.]]$
will give contributions with comparable strength.
Then, clearly the second term dominates at late times. Its 
contribution is found to be 
\begin{align}
 (m_{\rm 1-loop}^{\rm flux,s=1})^2  
 \approx -\frac {\tJ^2}{\rho^{2}\dim\mk }
 \Tr_{\mg}\Big( C^2[\mk]\,\frac{[\cF_{\ib\jb},[\cF^{\ib\jb},.]]}{\Box_6^2}\Big)
\ < \ 0 \,,
\end{align} 
which is significant and negative. 
A similar formula will also apply to $s>1$. Of course, the background is consistent iff the overall mass for all modes is positive or zero. This can be achieved by adding the positive mass contribution induced by the $\Box$ operator cf. \eqref{eq:m-kinetic-1-loop}, which applies to all spins $s=0,1,\ldots,\infty$. 

\subsection{Effective mass}
It is important to note that the above $m_{\oneloop}^{\kin}$ and $m_{\oneloop}^{\rm flux}$ are \emph{not} yet the physical mass, since the kinetic term is not yet appropriately normalized. With this in mind,
the effective mass of the lowest-spin fields
is therefore
\begin{align}
    (m_{\oneloop}^{R-\YM})^2 = (m_{\oneloop}^{\kin})^2(g_{\rm YM}^R)^2\,,
\end{align}
while the effective mass of their $\hs$ siblings read
\begin{align}
    (m_{\oneloop}^{R-\hs})^2 = \Big[(m_{\oneloop}^{\kin})^2+(m_{\oneloop}^{\rm flux})^2\Big](g_{\rm YM}^R)^2
\end{align}
which we assume to be small.
Assuming that the (positive) 1-loop contribution  \eqref{color-mass-bulk} dominates, the physical mass of ordinary YM gauge field is given by
\begin{align}
 (m_{\oneloop}^{R-\YM})^2 \approx \frac{1}{\frac{2}{\tJ g^2}\rho^2+\frac{\vartheta_{\mk}}{4(4\pi)^2}}\frac{\Delta_\cK^2}{\rho^2}\,.
\end{align}
Let us consider a similar scenario as in Fig. \ref{couplings}. Then, the behavior of $m^{R-\YM}_{\oneloop}$ in the early and late-time period of $\cM^{1,3}$ can be summarized as
\begin{table}[ht!]
    \centering
    \begin{tabular}{c||c}
      early period of $\cM^{1,3}$    &  late-time period of $\cM^{1,3}$ \\ \hline
      $(m_{\oneloop}^{R-\YM})^2\approx\frac{\Delta^2_{\cK}}{\rho^2\vartheta_{\mk}}$   & $(m_{\oneloop}^{R-\YM})^2\approx \frac{g\,\tJ\Delta_{\cK}^2}{\rho^4}$
    \end{tabular}
\end{table}\\
We conclude that 
all 
color $\msu(k)$ 
gauge bosons of $\hs$-IKKT acquire masses
in the presence of a large brane $\cK$,
with the mass scale set by the KK scales $(m_{\cK},\Delta_{\cK})$ 
of $\cK$.

\paragraph{Discussion.} In contrary to what one might hope, the $\mk\otimes\hs$-valued gauge fields do {\em not} seem to decouple at low energy. Furthermore,  the conventional lowest-spin gauge fields turn out to acquire a mass which is {\em larger} than the $\hs$-valued siblings, which is certainly \emph{not} desirable from a physics point of view.\footnote{There is another contribution from 
terms $\{\cF_{\dot\mu\dot\nu},y^\sigma\} \del_\sigma$, which comes from the non-commutative $\mmu(1)$-sector of the model that leads to
the induced Einstein-Hilbert term \cite{Steinacker:2021yxt}. However, we do not expect that this contribution will change the conclusion of our analysis, given that gravitational interaction is much weaker at low-energy regime.} 
This result 
suggests that the lowest-spin gauge fields are indeed massive, with 3 physical degrees of freedom\footnote{this is not in conflict with the result in \cite{Steinacker:2019awe} where massless fields were assumed for the physical Hilbert space, without quantum corrections. The no-ghost statement in \cite{Steinacker:2019awe} is expected to apply also in the massive case, due to the Yang-Mills form of the action and the physical state constraint.} as in Proca theory. Therefore, the low-energy limit of the $\hs$-YM theory considered in this work is {\em not} ordinary Yang-Mills theory.

\section{Scalar sector and stability of \texorpdfstring{$\cK$}{K}}\label{sec:5}

In this final section, we consider the one-loop contributions to the effective potential between disconnected sectors of the extra dimensions $\cK$, which are important to understand the stabilization of $\cK$ and possible Higgs potentials at low energy. We observe that $\cK$ with $|\cK|\leq 4$ does \emph{not} shrink to a point, thanks to an intricate interaction between $\cK$ and the $U(1)$ flux bundle over $\cM^{1,3}$.

\paragraph{Scalar potential at one loop.} The one-loop effective potential for the scalar sector can again be extracted from the general formula 
\eqref{eq:full-trace-HS-general}, where the background associated with $\cK$ is now unperturbed. Namely, we will turn off 
the Yang-Mills fields. As a result, we have
\begin{align}
    \chi^{(4)}_{\cM^{1,3}}[\alpha] = 4\,,\qquad \chi^{(2,\pm)}_{\cM^{1,3}}[\alpha] = 2
\end{align}
so that 
\small
\begin{align}\label{eq:full-trace-HS-scalar}
    \Gamma_{\oneloop}[\cK] 
    &= -\frac 1{2(2\pi)^4} 
 \!\!\int_{\cM^{1,3}} \!\! \sqrt{G} \rho^{-4}
 \int\limits_0^\infty \frac {d\a}{\a} \kappa[\a]\,
 \Tr_{\hs}\Big[  \Tr_{\mg\otimes V_{(6)}}\big(e^{-\im\a(\Box_6 +m_s^2  -\Sigma^{(6)}_{\ib\jb}[\cF^{\ib\jb},-])}\big)+
     \nn \\
     &\quad  + 2\Tr_{\mg}\big(
     e^{-\im\a(\Box_6+m_s^2)}\big)
     - \sum_\pm \Tr_{\mg\otimes \Psi_{(4\pm)}}\big(e^{-\im\a(\Box_6 +m_s^2 -\Sigma^{(4\pm)}_{\ib\jb}[\cF^{\ib\jb},-])}\big) \Big] \, \nn\\
     &=: -\frac 1{2(2\pi)^4} 
 \!\!\int_{\cM^{1,3}} \!\! \sqrt{G}\, V_\cK^{\oneloop}
\end{align}
\normalsize
thereby defining the effective potential $V_{\cK}^{\oneloop}$. 
This was considered as a function of the KK scale $m_\cK$ in \cite{Steinacker:2023myp} truncated at $\cO(\alpha^4)$, which leads to a stabilization of $\cK$. We will now study this expression more generally.
To proceed, we will again assume that the background $\cK$ is slowly varying such that $[\Box_6,\cF_{\ib\jb}] = 0$. Moreover, we can also approximate $m_s^2\approx 0$ since these IR masses scale as $\cO(R^{-2})$. Then, 
\small
\begin{align}\label{eq:full-trace-HS-scalar-UV}
V_{\cK}^{\oneloop} 
&\approx \tJ^2\rho^{-4}
 \int\limits_0^{\infty} \frac {d\a}{\a} \kappa[\a]\, \Tr_{\mg}\Bigg[e^{-\im\a\,\Box_6 } \Big(2+
\Tr_{V_{(6)}}\big(e^{\im\a\Sigma^{(6)}_{\ib\jb}[\cF^{\ib\jb},-]}\big)
     - \sum_\pm \Tr_{\Psi_{(4\pm)}}\big(e^{\im\a\Sigma^{(4\pm)}_{\ib\jb}[\cF^{\ib\jb},-]}\big)\Big) \Bigg] \, \nn\\
    &\equiv \tJ^2 \rho^{-4}
 \int\limits_0^{\infty} \frac {d\a}{\a} \kappa[\a]\, \Tr_{\mg}\Big(e^{-\im\a\Box_6} \sX^{(6)}[\cF]\Big)\,.
\end{align}
\normalsize
Here the trace over the $\hs$ modes just gives an overall factor $\tJ^2$, and 
$\sX^{(6)}[\cF]$ was defined in \eqref{eq:character-K-scalar-UV}.
Using the geometrical trace formula \eqref{trace-coherent-End}, the trace over $\mg$ can be evaluated as 
\begin{align}
\label{trace-character-prescale}
\Tr_{\mg}\Big(e^{-\im\a\Box_6} \sX^{(6)}[\cF]\Big)\,
 =  \int\limits_{\cK_x\times\cK_y}\! \frac{\Omega_x \Omega_y}{(2\pi)^{\frac{|\cK_x|+|\cK_y|}{2}}}\,
 e^{-\im\a(m^2_{\cK}(x-y)^2 + \Delta_\cK^2)}
 \sX^{(6)}[\delta\cF(x,y)]
\end{align}
where $\delta\cF(x,y) = \cF(x) - \cF(y)$ encodes the symplectic structure of $\cK$ evaluated at the two ends of the string. Note that we have rescaled $(x-y)^2\mapsto m_{\cK}^2(x-y)^2$ so that we only need to integrate over dimensionless variables. Therefore, \eqref{trace-character-prescale} becomes
\begin{align}
\label{trace-character}
\Tr_{\mg}\Big(e^{-\im\a\Box_6} \sX^{(6)}[\cF]\Big)\,
 = \int\limits_{\cK_x\times\cK_y}\! \frac{\Omega_x \Omega_y}{(2\pi)^{\frac{|\cK_x|+|\cK_y|}{2}}}\,
 e^{-\im\a(m_{\cK}^2(x-y)^2 + \Delta_\cK^2)}
 \sX^{(6)}[\delta\cF(x,y)]\,.
\end{align}
Let us now consider 
\begin{align}
\label{eq:character-K-scalar-UV-local}
 \sX^{(6)}[\delta\cF(x,y)] = 2+
\Tr_{V_{(6)}}\big(e^{\im\a\,\Sigma^{(6)}_{\ib\jb}\delta\cF^{\ib\jb}}\big)
 - \sum_\pm \Tr_{\Psi_{(4\pm)}}\big(e^{\im\a\,\Sigma^{(4\pm)}_{\ib\jb}\delta\cF^{\ib\jb}}\big) \, 
\end{align}
in more detail.
This formula applies 
for large branes $\cK$, which are well approximated by their semi-classical geometry.
We can now diagonalize $\delta \cF$
with a suitable $SO(6)$ rotation such that $\delta\cF$ is block-diagonal:
\small
\begin{align}
    \delta\cF^{\ib\jb}=\diag(\im \delta f_1\sigma_2\,,\im \delta f_2\sigma_2\,,\im \delta f_3\sigma_2)\,
    = \diag\Bigg[\begin{pmatrix}
        0 & \delta f_1 \\
        -\delta f_1 & 0
    \end{pmatrix},\begin{pmatrix}
        0 & \delta f_2 \\
        - \delta f_2 & 0
    \end{pmatrix},\begin{pmatrix}
        0 & \delta f_3 \\
        -\delta f_3 & 0
    \end{pmatrix}\Bigg]\,.
\end{align}
\normalsize
where $\sigma_2$ indicates a Pauli matrix. Choosing an appropriate basis for the $\gamma$-matrices as in \cite{Blaschke:2011qu}, we obtain  
\small
\begin{subequations}\label{eq:intermediate-characters}
    \begin{align}
    \sum_{\pm}\Tr_{\Psi_{(4\pm)}}\Big(e^{\im\alpha \Sigma_{\ib\jb}^{(4\pm)}\delta\cF^{\ib\jb}}\Big)&=(e^{\im\alpha \delta f_1}+e^{-\im\alpha \delta f_1})(e^{\im\alpha \delta f_2}+e^{-\im\alpha  \delta f_2})(e^{\im\alpha \delta f_3}+e^{-\im\alpha \delta f_3})\,,\\
    \Tr_{V_{(6)}}\Big(e^{\im \alpha\,\Sigma^{(6)}_{\ib\jb}\delta \cF^{\ib\jb}}\Big)&=e^{2\im\alpha \delta f_1}+e^{-2\im\alpha \delta f_1}+e^{2\im\alpha  \delta f_2}+e^{-2\im\alpha  \delta f_2}+e^{2\im\alpha  \delta f_3}+e^{-2\im\alpha \delta f_3}\,.
\end{align}
\end{subequations}
\normalsize
Intriguingly, we observe the following remarkable identity 
\small
\begin{align}\label{eq:character-chi-6-UV}
    \sX^{(6)}[\delta\cF] &=2\Big[1+\big[\cos(2\alpha \delta f_1)+\cos(2\alpha\delta f_2)+\cos(2\alpha\delta f_3)\big]-4 \cos(\a \delta f_1)\cos(\a\delta f_2)\cos (\a\delta f_3)\Big]  \nn\\
     &= -16\sin\Big[\frac\alpha 2(\delta f_1+\delta  f_2-\delta  f_3)\Big] \sin\Big[\frac\alpha 2(\delta f_1+\delta  f_3-\delta f_2)\Big] \times\nn\\
     &\qquad \qquad\qquad \qquad \times\sin\Big[\frac\alpha 2(-\delta f_1+\delta  f_2+\delta f_3)\Big] \sin\Big[\frac\alpha 2(\delta f_1+\delta f_2+\delta f_3)\Big]\, \nn\\
     &= -16 \sin\Big[\frac{\a}{2}\big(F(x)-F(y)\big)\Big]\prod_{i=1}^3 \sin\Big[\frac{\a}{2} \big(F(x) -F(y)- 2(f_i(x)-f_i(y))\,\big)\Big]
\end{align}
\normalsize
where $F(x):=f_1(x)+f_2(x)+f_3(x)$. This starts at order $\cO(\a^4)$
 in agreement with \eqref{eq:character-K-scalar-UV}, which reflects maximal SUSY. 
 We also notice that if the rank of $\cF$ is four (e.g. $f_3=0$), then this simplifies as
\small
\begin{align}
\label{chi6-rank4-full}
    \sX^{(6)}[\delta\cF]
    &=4(\cos(\alpha\delta f_1)-\cos(\alpha\delta f_2))^2= 
    2(\delta f_1^2-\delta f_2^2)^2
  \alpha^4+\cO(\alpha^6) \ \geq 0 \, ,
\end{align}
\normalsize
which is indeed positive (semi-)definite. 
We may write the leading $\cO(\a^4)$ term as
\begin{align}\label{eq:X6-character}
    \sX^{(6)}\approx\alpha^4\Big(4\Tr(\delta \cF^4)-(\Tr\delta \cF^2)^2\Big)=: \alpha^4\sQ^{(6)}[\delta \cF]
\end{align}
where 
\begin{align}
    \Tr\delta \cF^4\equiv\delta \cF^{\ib\jb}\delta \cF_{\jb\kb}\delta \cF^{\kb\mb}\delta \cF_{\mb\ib}\,,\qquad \Tr\delta \cF^2\equiv\delta \cF^{\ib\jb}\delta \cF_{\ib\jb}\,.
\end{align}
Carrying out the integral over $\alpha$,  the one-loop contribution to the effective potential\footnote{Note that this is the 1-loop effective potential in $\cN=4$ SYM for semi-classical background configurations.} is obtained as
\small
\begin{align}\label{eq:potential-for-K}
    V^{\oneloop}_{\cK} 
    &=-\frac{\tJ^2\pi^2}{ \rho^4}
 \int\limits_{\cK_x\times \cK_y}\! \frac{\Omega_x\Omega_y}{(2\pi)^{\frac{|\cK_x|+|\cK_y|}{2}}}\,\frac{4\Tr(\cF(x)-\cF(y))^4 -(\Tr(\cF(x)-\cF(y))^2)^2}{(m_{\cK}^2(x-y)^2+\Delta_{\cK}^2)^2}\,.
\end{align}
\normalsize
Note that 
the above integrals over the positions do \emph{not} localize for branes with dimension larger or equal to 4, i.e. they range over the entire brane $\cK_x \times \cK_y$. However, they are always finite since $\cK$ is compact\footnote{
We recall that 
$\cF$ has dimension $[mass^2]$ while $x,y$ are dimensionless. The symplectic volume integrals themselves are dimensionless numbers, cf. \eqref{Bohr-Sommerfeld}. Therefore, the above potential has dimension $[mass^4]$, as it should be.}. 
For smaller branes such as point branes, the above trace needs to be evaluated explicitly; this will be discussed below.

For illustrative purpose, let us consider a few cases where $\sX^{(6)}$ has a simple form. 
\begin{itemize}

\item If $\delta \cF$ has rank-2, then $\sX^{(6)} > 0$ is positive, leading to a negative potential. 
However, it is expected that this case is too degenerate.

 \item In the case where $\delta f_3=0$ or $\delta\cF$ is rank-4, $\sX^{(6)}=4(\cos(\alpha \delta f_1)-\cos(\alpha \delta f_2))^2\geq 0$ is again positive, leading to a negative potential and suggesting a bound state. This case will be discussed in more detail below.
    \begin{figure}[ht!]
        \centering
        \includegraphics[scale=0.38]{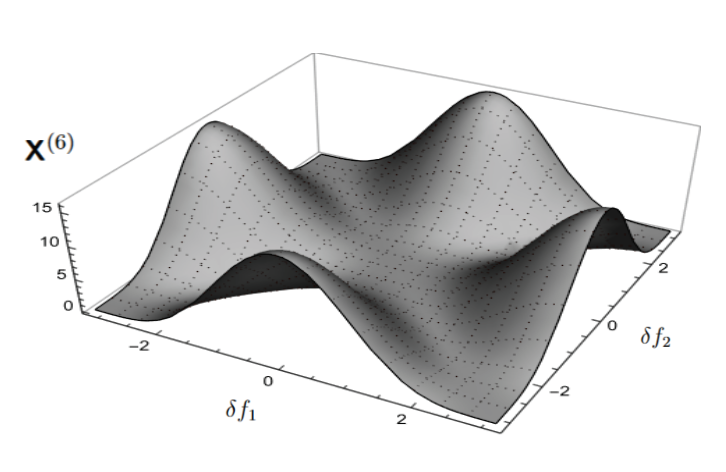}
        \caption{A generic form for $\sX^{(6)}$ if $\delta \cF$ is rank-4.}
        \label{fig:potential4d}
    \end{figure}
    
    \item Finally, consider the case where $\delta \cF$ is rank-6. For simplicity, suppose $\delta f_1\approx \delta f_2 \approx \delta f_3$. Then, $\sX^{(6)}[\cF]\Big|_{\delta f_i=\delta f}=-16 \sin\Big(\frac{\alpha \delta f}{2}\Big)^3\sin\Big(\frac{3\alpha \delta f}{2}\Big)$ and has the form of Fig \ref{fig.rank-6}. 
    \begin{figure}[ht!]
        \centering
        \includegraphics[scale=0.33]{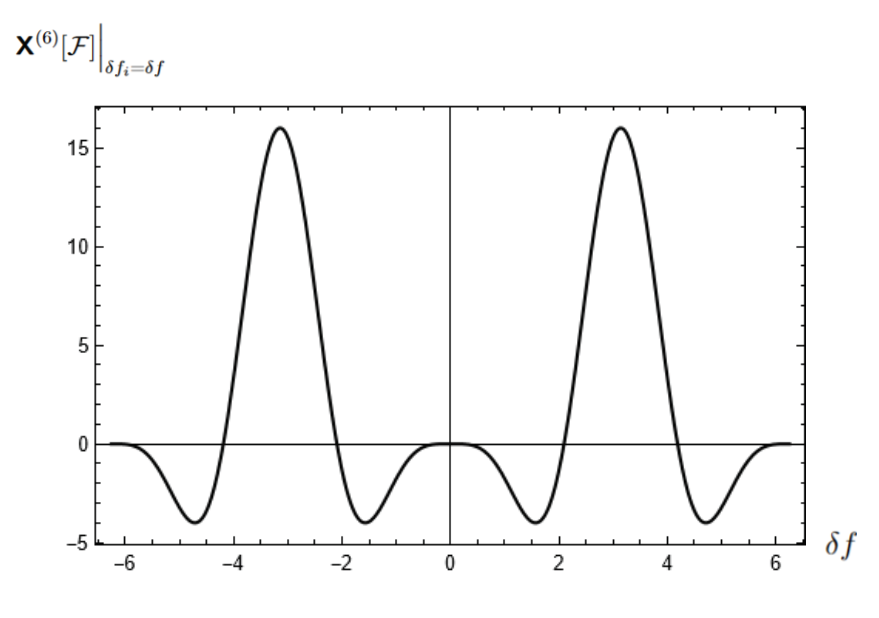}
        \caption{$\sX^{(6)}[\cF]$ where $\delta f_i=\delta f$}
        \label{fig.rank-6}
    \end{figure}
   In this case, some finite repulsive force between branes  may arise due to $\sX^{(6)}[\delta f]<0$. 
\end{itemize}



\subsection{Examples of scalar potentials}
\paragraph{Point branes.} For point branes with $\cH_{\cK_i}\simeq \C$, we simply have $\cF_{\ib\jb}=0$, thus $V_{\cK}^{\oneloop}=0$. 
This is to be expected, because the theory locally reduces to (a $\hs$ extension of)  $\cN=4$ SYM. 


\paragraph{A point brane and a bulk brane.}Consider a point brane $\cK_{1}^{(x)}$ located at $y$ and a large brane $\cK$ centered at the origin. Then their interaction is mediated by the integral over string modes stretching between these branes. The effective potential is  
\begin{align}
\label{point-bulk-interaction}
    V_{\cK}^{\oneloop}[\cK_1^{(x)},\cK] 
    =-\frac{\tJ^2}{ \rho^4}
 \int\limits_{\cK}\! \frac{\Omega_y}{(2\pi)^{|\cK|/2}}\,\frac{4(\Tr\cF^4)-(\Tr(\cF^2))^2}{(m_{\cK}^2(x-y)^2+\Delta_{\cK}^2)^2}\,,\qquad \cF=\cF(y)\,.
\end{align}
This gives the effective potential $V_1(x)$ for the point brane. Observe that the potential has a non-trivial extremum at $x=0$. Furthermore, since $V_{\cK}^{\oneloop}[\cK_1^{(x)},\cK]<0$, the point branes tend to glue to $\cK$, which should lead to a color-like low-energy gauge sector. 
This means that the point branes are attracted to the (center of) $\cK$, as desired.
In fact, this will be the standard behavior of branes with dimensions lesser or equal to four.

Let us evaluate the above effective potential quantitatively by recalling that $\cF = \cO(\Delta_\cK^2)$. Then, for
\begin{subequations}\label{eq:VUV-point-vs-bulk}
    \begin{align}
        |\cK|&=2\,,\qquad &V_{\cK}^{\oneloop}[\cK_1^{(0)},\cK]&\propto -\frac{\tJ^2\,d_{\cK}\Delta^6_{\cK}}{\rho^2(m_{\cK}^2+\Delta_{\cK}^2)}\approx -\frac{\tJ^2\Delta_{\cK}^4}{\rho^2}\,,\\ 
        |\cK|&=4\,, \qquad &V_{\cK}^{\oneloop}[\cK_1^{(0)},\cK]&\propto 
        -\frac{\tJ^2\Delta_{\cK}^4}{\rho^2}\log(1+d_{\cK}^{1/2})\,,\\ 
        |\cK|&=6\,, \qquad &V_{\cK}^{\oneloop}[\cK_1^{(0)},\cK]&\propto \pm \frac{\tJ^2\Delta^4_{\cK}}{\rho^2}\Big(d_{\cK}^{1/3}-2\log(1+d_{\cK}^{1/3})\Big) \,,
    \end{align}
\end{subequations}
for large $d_{\cK}$. Note that since $\delta\cF$ is rank-6 when $|\cK|=6$, the last expression can have either sign. As a reminder, the curvature radius $R_{\cK} = m_\cK =d_{\cK}^{1/|\cK|}\Delta_{\cK}$ encodes the size of $\cK$. 

\paragraph{Intersecting flat branes.}


Consider the case of two almost-flat large branes $\cK_1$ and $\cK_2$, which may intersect each other.
Then the interaction between the branes is obtained from \eqref{eq:potential-for-K} assuming that $\cF_{1,2} \approx const$ for each brane:
\begin{align}\label{eq:flat-branes-VUV}
    V_{\cK}^{\oneloop}[\cK_1,\cK_2] 
    &\propto -
 \int\limits_{\cK_1\times \cK_2}\! \frac{\Omega_x\Omega_y}{(2\pi)^{\frac{|\cK_1|+|\cK_2|}{2}}}\,\frac{4\Tr(\cF_1-\cF_2)^4 -(\Tr(\cF_1-\cF_2)^2)^2}{(m_{\cK}^2(x-y)^2+\Delta_{\cK}^2)^2}\,.
\end{align}
\normalsize
 Note again that $\cK_i$ are assumed to be compact\footnote{The integrals would be divergent for non-compact branes, so that the present result should only be used to gain a qualitative understanding.}. 
 The above integral is always negative or zero 
as long as ${\rm rank}(\delta \cF) \leq 4$ due to \eqref{chi6-rank4-full}, and vanishes only if $\delta \cF = \cF_1 - \cF_2$ vanishes or is (anti-) self-dual\footnote{Namely, $\delta \cF = \pm \ast \delta \cF$ in the 4-dimensional case with $\ast$ being the Hodge dual operator on $\cK_i$. Indeed, one can show that $4\Tr(\cF^4)-(\Tr(\cF^2))^2=(\cF+\ast \cF)^2(\cF-\ast \cF)^2$, which justifies the claim.}. 

If one of the branes has dimension $6$, then  $V_{\cK}^{\oneloop}[\cK_1,\cK_2]$ can have either sign. Indeed, 
consider first the numerator 
 $\omega[\delta \cF] :=4\Tr(\cF_1-\cF_2)^4-(\Tr(\cF_1-\cF_2)^2)^2$.
We can use \eqref{eq:character-chi-6-UV} to write
\begin{align}\label{stability-flat-brane}
   \omega[\delta \cF]= -(\delta f_2+\delta f_3-\delta f_1 ) (\delta f_1 + \delta f_2 - \delta f_3) (\delta f_1 - \delta f_2 + \delta f_3) (\delta f_1 + \delta f_2 + \delta f_3)\,.
\end{align}
For the sake of argument, 
assume that $\cF_1$ has rank $6$ and $\cF_2$ has rank $4$, so that we can replace $\delta f_1 \to f_1$. Then, clearly $\omega[\delta \cF]$ -- and hence the overall interaction energy -- can be either positive or negative.
In particular, if $f_1> 2\max\{|\delta f_2|,|\delta f_3|\}$, the potential is always positive.

Let us estimate \eqref{eq:flat-branes-VUV} in various cases where $|\cK_{1,2}|\leq 4$. 
We assume for simplicity that the branes are orthogonal, so that $(x-y)^2 = x^2 + y^2$ and $|\cK_1|+|\cK_2|\leq 6$. Then, for
\begin{align}\label{flat:d<6}
    |\cK_1|&=2\,,|\cK_2|=2\,,\qquad V_{\cK}^{\oneloop}\propto -\frac{m_{\cK}^4}{d_{\cK}^2}\log\Big[\frac{(1+d_{\cK})^2}{1+2d_{\cK}}\Big]\approx -\frac{m^4_{\cK}}{d^2_{\cK}}\log(d_{\cK})\approx -\Delta^4_{\cK}\log(d_{\cK})<0\,.
\end{align}
assuming $\cF_{1,2}\approx const $ of order $\cO(\Delta_{\cK}^2)$ and large $d_{\cK_x}\approx d_{\cK_y}$. Observe that the integral over $x$ and $y$ is actually dominated by the long strings. Thus, the above result is only indicative for understanding the relevant scales\footnote{In particular, this means that the contributions from (almost-) zero modes arising at the intersection are not significant and can be dropped here.}.
Nevertheless, the net result is always finite due to the compact nature of $\cK$. Alternatively, the above computation can be carried out in terms of KK modes on $\cK$, see e.g. \cite{Steinacker:2023myp} for more details. However, the present geometric evaluation is much more transparent.

\paragraph{Self-intersecting branes.} The above discussion applies, 
in particular, for self-intersecting branes such as squashed $\P^2_*$, cf. \cite{Steinacker:2015dra}, which are interesting due to chiral fermions  arising at their intersection.
\begin{figure}[ht!]
    \centering
    \includegraphics[scale=0.48]{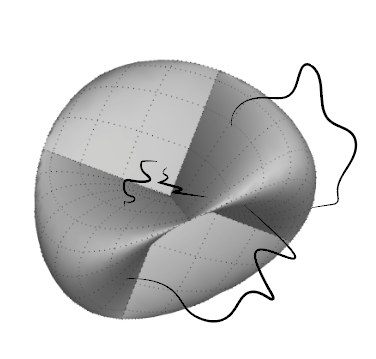}
    \caption{A squashed self-intersecting brane $\P^2_*$  with would-be zero modes as short open strings near the intersection of different sheets of $\P^2_*$, as well as long strings contributing in the UV.}
\end{figure}\\
For example, the $4d$ self-intersecting sheets of 
$\P^2_*$ can be approximated by a system of flat intersecting branes sharing a $2d$-plane, which may be written as $\R^4_{4567} \cap \R^4_{4589}=\R^2_{45}$. The open string modes between two sheets then lead to
\begin{align}
\label{eq:VUV-self-intersecting-K}
    V_{\cK}^{\oneloop}[\cK_1,\cK_2]\propto -\int_{\cK_1\cap \cK_2}\frac{\Omega_1\Omega_2}{(2\pi)^{\frac{|\cK_1|+|\cK_2|}{2}}}\frac{4\Tr(\cF_1-\cF_2)^4-(\Tr(\cF_1-\cF_2)^2)^2}{(m^2_{\cK}x^2+m^2_{\cK}y^2+m^2_{\cK}(z_1-z_2)^2+\Delta^2_{\cK})^2} \ < 0
\end{align}
which is expected to be negative since 
the flux difference $\delta \cF$ on the intersection $\R^2_{45}$ vanishes, so that
$\delta \cF$ has  (essentially) rank 4. At the end of the day, the potential between different sheets will again be proportional to $m_\cK^4$, since this is the unique UV scale of $\cK$. 

Note that the contributions from possible zero modes arising on self-intersecting branes are of order 
$\cO(\Delta_{\cK}^4)$
and can therefore be neglected. This contribution can be estimated quickly by integrating over the radial direction of a sphere of radius $\Delta_{\cK}$ with the intersection of the self-intersecting brane as its origin. 





\subsection{\texorpdfstring{$U(1)$}{U1}-flux bundle contributions to the scalar potential} 

As pointed out in \cite{Steinacker:2024}, there is another contribution to the vacuum energy on the background under consideration, which
arises from the geometric background  $\cF^{\dot\mu\dot\nu}$ on $\cM^{1,3}\times S^2_{\tJ}$ coupling to $\cK$ at one loop. This leads to  stabilization of $\cK$,
due to the non-commutativity scale $\Delta_{\tJ}$ of $S^2_{\tJ}$. Note that $\Delta_{\tJ}$ can be related to the scale  $L_{\rm NC}$ of $\cM^{1,3}$, cf. Subsection \ref{sec:K-scales}. This represents a crucial difference to ordinary $\cN=4$ SYM, which does \emph{not} have any scale.

To compute this contribution, we need to take into account the $U(1)$-valued background flux $\cF^{\dot\mu\dot\nu}$ on space-time in 
\eqref{eq:full-trace-HS-general} where instead of $3+1$ dimensional character $\chi$ cf. \eqref{vector-Gamma-1-loop}, we will consider the $9+1$ dimensional one and expand it to order $\cO(\cF^4)$. 
Then the 1-loop potential \eqref{eq:full-trace-HS-scalar-UV} takes the form
\small
\begin{align}
V^{\oneloop}_{\cM\times S^2_{\tJ}\times\cK} &=  \rho^{-4}
 \int\limits_0^\infty \frac{d\a}{\a}\,\kappa[\a] \alpha^4\,\Tr_{\hs\otimes\mg}
 e^{-\im\a (\Box_6 + \Delta_\cK^2)}
\sQ[\delta \cF]= - \frac{\pi^2}{\rho^4}
\Tr_{\hs\otimes\mg}\Big[\frac {\sQ[\delta \cF]}{(\Box_6 + \Delta_\cK^2)^2}
 \Big]
\end{align}
\normalsize
where $\delta \cF = [\cF_{\Ibold\Jbold},.]$ and $\sQ[\delta \cF] = 4 \delta \cF^{\Ibold\Jbold}\delta \cF_{\Jbold\Kbold}\delta \cF^{\Kbold\Lbold}\delta \cF_{\Lbold\Ibold}
    - (\delta \cF^{\Ibold\Jbold}\delta \cF_{\Ibold\Jbold})^2$ with $\Ibold=0,1,\ldots,9$.

\begin{figure}[ht!]
    \centering
    \includegraphics[scale=0.37]{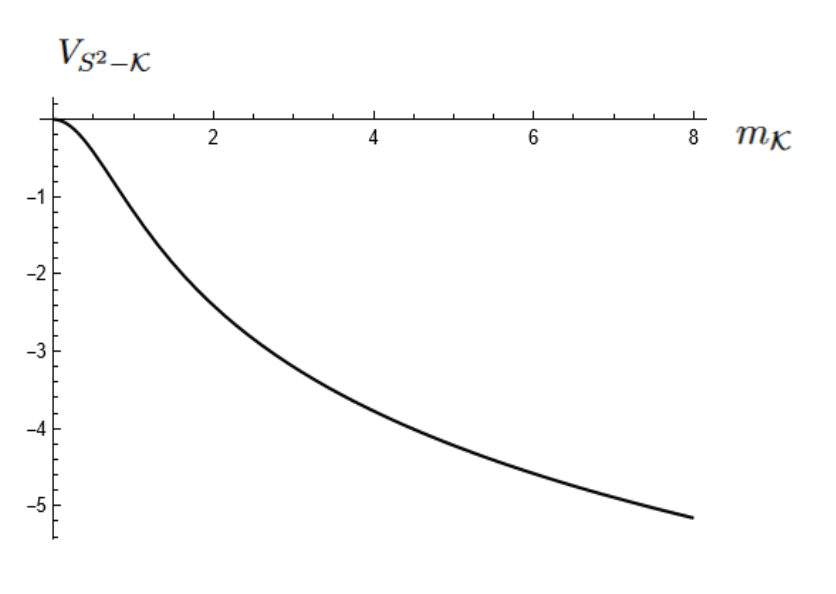}
    \caption{Interaction potential between $\cK$ and $S^2$.}
    \label{fig:V-S2-K}
\end{figure}

\paragraph{$S^2-\cK$ contribution.} Now consider the  mixed $S^2_\tJ -\cK$ term, where 
\begin{align}
     \sQ[\delta\cF]\Big|_{S^2_\tJ -\cK}=+(\delta\cF^{0i}\delta \cF^{0i})(\cF_{\ib\jb}\cF^{\ib\jb})\,,\qquad i=1,2,3,
\end{align}
since $\cF^{\dot\mu \ib} = 0$ as assumed above\footnote{Note that the space-like contributions $-(\delta\cF^{ij}\delta \cF^{ij})(\cF_{\ib\jb}\cF^{\ib\jb})$ to $\sQ[\delta\cF]$ are strongly suppressed for $\cM^{1,3}$ and hence omitted.
This would typically not be the case for the Moyal-Weyl quantum plane $\R^4_{\theta}$, which provides a dynamical mechanism in favor of $\cM^{1,3}$.}. 
The trace over $\mg$ can be evaluated as always using the geometric trace formula for $\cK$. The trace over $\hs$ can also be evaluated using string modes,
noting that $\delta \cF^{\dot\mu\dot\nu} \sim \cF^{\dot\mu\dot\nu}(u) - \cF^{\dot\mu\dot\nu}(v)$ when 
acting on a string mode $|u\rangle\langle v|$ on $S^2_\tJ$.
Recalling that $\cF^{0i} \sim \cosh(\tau) \ell_p^2\, u^i$ where $r$ is the intrinsic length scale of $S^2_\tJ$ and $u^i$ are the normalized fiber coordinates of $S^2_{\tJ}$, cf. Appendix \ref{app:A}.
We obtain
\small
\begin{align}
V_{S^2-\cK}^{\oneloop} &= -\rho^{-4} \int\limits_{S^2 \times S^2}\frac{\varpi_u \varpi_v}{(2\pi)^2}
 (\delta\cF^{0i}\delta \cF^{0i})
\int\limits_{\cK \times \cK} 
\frac{\Omega_x \Omega_y}{(2\pi)^{\frac{|\cK_x|+|\cK_y|}{2}}}\,
\frac{\cF_{\ib\jb}\cF^{\ib\jb}}{[m_\cK^2(x-y)^2 + \Delta_\cK^2 + \Delta_{\tJ}^2]^2}\,
\end{align}
\normalsize
where $\varpi$ is the symplectic volume form of $S^2_{\tJ}$ \eqref{S2-volume-form}. The integral over $\cK$ results in
\small
\begin{align}\label{eq:IK}
   I_\cK\equiv-\int\limits_{\cK \times \cK} 
\frac{\Omega_x \Omega_y}{(2\pi)^{\frac{|\cK_x|+|\cK_y|}{2}}}\,
\frac{1}{[m_\cK^2(x-y)^2 + \Delta_\cK^2 + \Delta_{\tJ}^2]^2}  \approx\frac{d^2_{\cK}}{m^4_{\cK}}\Bigg[\frac{m_\cK^2}{m_\cK^2+\Delta_\tJ^2}-\log\Big(\frac{m^2_{\cK}+\Delta^2_{\tJ}}{\Delta^2_{\tJ}}\Big)\Bigg]
\end{align}
\normalsize
for large $\cK$. Here, $\Delta_{\tJ}$ 
is the scale of non-commutativity on $S^2_\tJ$,
which is 
approximately $\frac 1{\ell_p^2}$. On the other hand,
\begin{align}
    I_{\cF^2_{\cM}}\equiv\int \limits_{S^2 \times S^2}\frac{\varpi_u \varpi_v}{(2\pi)^2}\,(\delta\cF^{0i}\delta \cF^{0i})
    \approx r^{4}\tJ^2 \cosh^2(\tau) 
    \sim  r^{4}\tJ^2 \rho^{4/3}\,,\qquad i=1,2,3\,,
\end{align}
As a result,
\small
\begin{align}
V_{S^2-\cK}^{\oneloop} &= \rho^{-4} I_{\cF^2_{\cM}}\times I_{\cK}\, \Delta_{\cK}^4 
\approx \frac{r^4 \tJ^2d_{\cK}}{\rho^{8/3}}
\Bigg[\frac{m_\cK^2}{m_\cK^2+\Delta_\tJ^2}-\log\Big(\frac{m^2_{\cK}+\Delta^2_{\tJ}}{\Delta^2_{\tJ}}\Big)\Bigg] \ < 0\,,
\end{align}
\normalsize
using $\cF_{\ib\jb} = \cO(\Delta_\cK^2)$ and $\Delta_\cK^4 = m_\cK^4/d_\cK$, which provides the crucial negative slope required for a non-trivial vacuum, as elaborated below. Here, we have
assumed that $d_\cK \gg 1$ and replace $\cK$ with a 4-dimensional ball 
with radius of $\cO(1)$. This potential behaves as 
\begin{align}
    V_{S^2-\cK}^{\oneloop} \sim - \frac{r^4\tJ^2 d_{\cK}}{2\Delta_\tJ^4} \rho^{-8/3} m_\cK^4
\end{align}
near 0, while for large $m_\cK$ it has a mild $V_{S^2-\cK}^{\oneloop} \sim -\log(m_\cK)$ behavior as shown in figure \ref{fig:V-S2-K}.

Combined with $V_0 + V_\cK^{\oneloop} \sim m_\cK^4$, this leads to a non-trivial minimum for $m_\cK$ as discussed in the next section, given that the downward slope of
$V_{S^2-\cK}^{\oneloop}$ dominates at small $m_\cK$. The crucial point we want to emphasize here is that the coupling between $\cK$ and the $U(1)$ flux on space-time introduces a scale into the theory.
This qualitative conclusion is unchanged by the further effects from $\cK$ as discussed below.

It is interesting to note that the induced Einstein-Hilbert term also arises from the mixed interaction between $\cK$ and $\cM^{1,3}$, taking into account derivative contributions of the background cf. \cite{Steinacker:2021yxt,Steinacker:2023myp}.

\paragraph{$S^2_{\tJ}$ contribution.}

Finally, consider the contribution from $\cO(\cF_{\dot\mu\dot\nu}^4)$, which is given by 
\small
\begin{align}
V_{S^2}^{\oneloop} &= -\rho^{-4} \int\limits_{S^2 \times S^2}\frac{\varpi_u \varpi_v}{(2\pi)^2}\,\sQ[\delta \cF_{\cM}]
\int\limits_{\cK \times \cK} 
\frac{\Omega_x \Omega_y}{(2\pi)^{\frac{|\cK_x|+|\cK_y|}{2}}}\,
\frac{1}{[m_\cK^2(x-y)^2 + \Delta_\cK^2 + \Delta_{\tJ}^2]^2}  
\end{align}
\normalsize
where $\sQ[\delta \cF_\cM]:=\sQ[\delta \cF]\big|_{S^2_{\tJ}}$. Using \eqref{eq:IK}, we get
\begin{align}
    V^{\oneloop}_{S^2}=\rho^{-4} I_{\cF^4} \times I_{\cK}\,,\qquad I_{\cF^4}\equiv \int\limits_{S^2 \times S^2}\frac{\varpi_u \varpi_v}{(2\pi)^2}\,\sQ[\delta \cF_{\cM}]\,.
\end{align}
We can evaluate the tensorial contribution again
using string modes on $S^2_{\tJ}$
as  
\begin{align}
    \delta \cF^{\dot\mu\dot\nu}\delta \cF_{\dot\nu\dot\rho}\delta \cF^{\dot\rho\dot\sigma}\delta \cF_{\dot\sigma\dot\mu}
  %
= 2 (\delta \cF^{0j}\delta \cF_{j0})^2
    + 4\delta \cF^{0j}\delta \cF_{jk}\delta \cF^{kl}\delta \cF_{l 0} 
+  \delta \cF^{ik}\delta \cF_{kl}\delta \cF^{ln}\delta \cF_{n i} 
\end{align}
where the purely space-like contributions are subleading. Then,
\begin{align}
    \sQ[\delta \cF_{\cM}] \approx 4(\delta \cF^{0j}\delta \cF_{j0})^2 >0
\end{align}
for long string modes. We obtain
\begin{align}
    \int \limits_{S^2 \times S^2}\frac{\varpi_u \varpi_v}{(2\pi)^2}\,\sQ[\delta \cF_{\cM}]
    \approx 4r^{8}\tJ^2 \cosh^4(\tau) \,.
\end{align}
In terms of the magnitude, we observe that 
\begin{align}
    V_{S^2}^{\oneloop} \sim \frac{\Delta_\tJ^4}{\Delta_\cK^4} V_{S^2-\cK}^{\oneloop} \ll V_{S^2-\cK}^{\oneloop}\,,\qquad \frac {\p}{\p m_\cK}V_{S^2}^{\oneloop} > 0\,.
\end{align}
The bottom line of this computation is that even though $V_{S^2}^{\oneloop}<0$ does
contribute a binding energy\footnote{It was assumed in \cite{Steinacker:2024,Kumar:2023bxg} that $V_{S^2}^{\oneloop} >0$, which suggested a different mechanism for stabilizing $\cK$. The present result using geometric trace formula seems more satisfactory.}, it only helps to stabilize the $\cM^{1,3}$ background, but not $\cK$.

\subsection{Stability of \texorpdfstring{$\cK$}{K}}

Given the above results, we can try to address the important problem of how to stabilize $\cK$. It is difficult to answer this question in full detail; thus we will merely try to understand some semi-quantitative features, and leave the complete analysis for future work. 

The stability of  $\cK$ is determined by the condition that the 1-loop effective potential
\begin{align}
\label{V-tot}
    V_{\rm eff}[\cK]
    &\approx   \V_{0}[\cK]  +
    V^{\oneloop}_{\cK}[\cK]+V^{\oneloop}_{S^2}+V^{\oneloop}_{S^2-\cK}
\end{align}
should have a non-trivial (global or local) minimum, which is negative. Here, 
\begin{align}\label{eq:YM-K}
    V_{0}[\cK] = \frac {\tJ}{g^2\rho^2}\cF^{\ib\jb} \cF_{\ib\jb}\approx \frac{\tJ \Delta_{\cK}^4}{g^2\rho^2}\approx \frac{\tJ}{g^2\rho^2}\frac{m^4_{\cK}}{d_{\cK}} >0
\end{align}
Note that around $m_{\cK}=0$, $V_{tot}\sim V_{S^2}\sim -\frac{1}{4\Delta_{\tJ}\rho^4}$. 
Then, to ensure global stability, we will assume that the classical potential dominates 
over the 1-loop potentials for large $m_\cK$, i.e. $V_0[\cK]>\eqref{eq:VUV-self-intersecting-K}$, which is a very reasonable requirement in the weak coupling regime under consideration. The non-trivial local minimum for $m_\cK > 0$ arises from the negative contribution in $V_{S^2-\cK}^{\oneloop}$. In summary, $V_{\rm eff}[\cK]$ has the following shape:
\begin{figure}[ht!]
    \centering
    \includegraphics[scale=0.45]{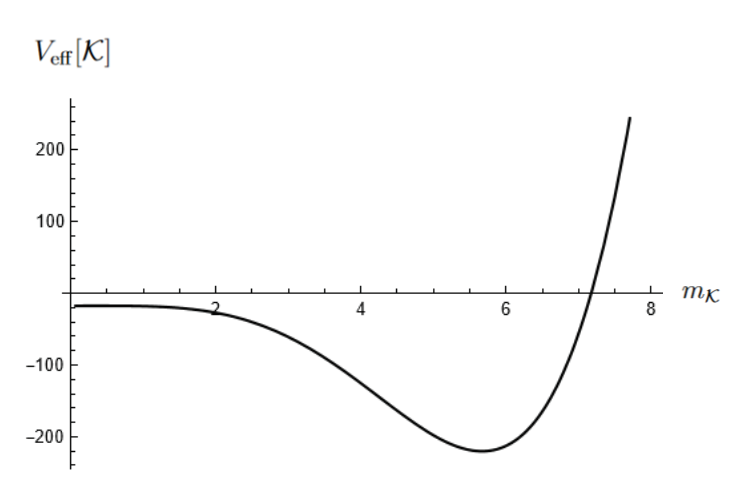}
    \caption{The potential $V_{\rm eff}$ with a global minimum for some finite value of $m_{\cK}>0$.  
}
    \label{Fig:Vtot}
\end{figure}\\
The reader may notice that the structure of $\cK$ is largely determined by the non-abelian sector of the theory, while the coupling to $S^2_{\tJ}$ is essential for stabilizing the radial scale. Furthermore, the interaction potential $V_{\cK_i-\cK_j}$ between two branes $\cK_i\subset \cK$ is 
attractive, provided the 
``effective" embedding dimensions $|\cK_i|$ less than or equal to 4. 
This strongly suggests the existence of stable branes $\cK$ with non-trivial sub-structure, as considered in this paper. Determining their specific structures would require much more detailed work.

\section{Discussion}\label{sec:discussion}

This work contains several significant new results, which should provide the basis for further investigations of the physics in the matrix model framework.
The first important (but rather disappointing) result is the mechanism described in Section \ref{sec:massive}, which induces mass for all (unbroken) $\hs$-valued gauge fields from quantum effects in the presence of 
$\cK$. While this mechanism allows $\hs$-YM to avoid no-go theorems \cite{Weinberg:1964ew,Coleman:1967ad}, it does not lead to near-realistic physics\footnote{One possibility to circumvent this issue is to consider the minimal quantum space with $\tJ=0$. This will be elaborated elsewhere.}. Note, however, that the time-dependent effective couplings of $\hs$-YM at one loop do show reasonable behavior (cf. Fig. \ref{couplings}).



A further  important result is the scalar potential at one loop, which allows for the stabilization of $\cK$. This  is also essential for obtaining an induced Einstein-Hilbert term and gravity in the present framework, along the lines of \cite{Steinacker:2021yxt,Steinacker:2023myp,Kumar:2023bxg}. 
It turns out that the scalar potential indeed acquires 
a non-trivial minimum for a non-vanishing radius of suitable $\cK$, with scale inherited from the non-commutativity scale of the 
cosmological space-time $\cM^{1,3}$. 
The methods developed here provide the basis to determine the dynamically preferred structure of the compact extra dimensions $\cK$, which should be elaborated in future work.
The crucial point we want to emphasize here is that the superconformal symmetry of the $\cN=4$ SYM sector in $\hs$-YM is broken by 
 the quantum spacetime. This provides an essential ingredient for obtaining interesting physics.

Finally, we have computed the 1-loop corrections to the Yang-Mills coupling constants. All of these calculations rely crucially on a basic trace formula based on coherent states, which provides a suitable way to perform loop computations, despite the complexity of the model under consideration.
These techniques will certainly be useful also in other contexts, and provide an important step towards identifying more realistic backgrounds in the matrix model.


Some further comments are in order, to clarify the relation of the present work with other approaches. The UV finiteness of the higher-spin gauge theory induced by the IKKT matrix model 
on the background brane $\cM^{1,3}\times S^2_{\tJ}\times \cK$ is essential for the loop computations, based on to maximally supersymmetry and the fuzzy nature of $\cK$. This is quite different with standard massless higher-spin gravities, see e.g. \cite{Giombi:2013fka,Giombi:2014iua,Giombi:2014yra,Beccaria:2015vaa,Beccaria:2014jxa,Bae:2016rgm,Gunaydin:2016amv} where one typically needs to provide some regularization scheme. The present matrix model setting also differs from the conventional approach to string theory, where 
target space is compactified; this is {\em not} the case here.
Furthermore, the mechanism of induced masses found in this work should allow obtaining a complete effective theory for massive higher-spin fields 
where all vertices can be derived by averaging over $S^2_{\tJ}$. It would also be interesting to compare  interacting vertices of $\hs$-IKKT with recent attempts of constructing Lorentz-invariant vertices of massive fields in e.g. \cite{Arkani-Hamed:2017jhn,Metsaev:2022yvb,Cangemi:2023ysz}. 


\section*{Acknowledgement}
We thank Tim Adamo and Zhenya Skvortsov for enlightening discussion. TT thanks the Simons Foundation for the hospitality during the annual meeting on Celestial Holography in Newyork, where the final stage of this work was in completion. This work is supported by the Austrian Science Fund (FWF) grant P36479.

\appendix

\section{Supplemental material}
\label{app:A}

\subsection{Useful relations}
This appendix provides additional relations extracted from \cite{Sperling:2018xrm,Steinacker:2019awe,Sperling:2019xar} for $\cM^{1,3}$ used in the main text. We have
\begin{subequations}
\begin{align}
    y_{\mu}\ttb^{\mu}&=0\,,\qquad \qquad \qquad \qquad \quad \quad \qquad \qquad \qquad\quad  \  \mu=0,1,2,3\,, \label{eq:orthogonalofPY}\\
    \eta_{\mu\nu}\ttb^{\mu}\ttb^{\nu}&=\frac{1}{\ell_p^2}+\frac{y_4^2}{\ell_p^2R^2}=+\ell_p^{-2}\cosh^2(\tau)\,, \qquad \qquad \quad \ \eta_{\mu\nu}=\diag(-,+,+,+)\,, \label{S2sphereM13}\\
   y^\mu y_\mu &= - R^2 \cosh^2(\tau) \,,
   \label{yy-square} \\ 
     \{\ttb^{\mu},y^{\nu}\}&=+\frac{\eta^{\mu\nu}}{R}y^4=\eta^{\mu\nu}\sinh(\tau)\,,\\
    \{\ttb^{\mu},y^4\}&=-\frac{y^{\mu}}{R}\,,\\
    m^{\mu\nu}&=-\frac{\theta^{\mu\nu}}{\ell_p^2}=-\frac{1}{\cosh^2(\tau)}\Big(\sinh(\tau)(y^{\mu}\ttb^{\nu}-y^{\nu}\ttb^{\mu})+\eps^{\mu\nu\sigma\rho}y_{\sigma}\ttb_{\rho}\Big)\label{mgenerator}\,,
\end{align}
\end{subequations}
which are frequently used in the main text. 
Here $m^{\mu\nu}$ are the generators of $\mso(1,3) \subset \mso(2,4)$ acting on $\cH$, and $\ell_p = \frac 2\tJ R$ is an intrinsic length scale on $\cM^{1,3}$. At late times, we have 
\begin{align}
\label{theta-munu-approx}
    \theta^{\mu\nu}\approx \frac{\ell^2_p}{\cosh(\tau)}(y^{\mu}\ttb^{\nu}-y^{\nu}\ttb^{\mu})
\end{align}
which is dominated by the space-time 
components $\theta^{0i}$ of size 
\begin{align}
    |\theta^{0i}| \sim L_{\rm NC}^2 \sim R\, \ell_p \cosh(\tau)
\end{align}
It is useful to introduce normalized generators 
\begin{align}
u^{\mu}=\frac{\ell_p}{\cosh(\tau)}\ttb^{\mu}\, \qquad\text{such that}\qquad  u_\mu u^\mu = 1 \ .
\end{align}
Then 
\begin{align}\label{flat-theta}
 \theta^{\mu\nu}\approx \frac{\ell_p\sinh(\tau)}{\cosh(\tau)}(y^{\mu}u^{\nu}-y^{\nu}u^{\mu})\approx \ell_p(y^{\mu}u^{\nu}-y^{\nu}u^{\mu})\,. 
\end{align}

\subsection{The structure of the higher-spin sphere \texorpdfstring{$S^2_\tJ$}{S2}}\label{sec:S2J-structure}

At some given reference point $\tp\in \cM^{1,3}$ where $y^\mu\big|_{\tp} = (y^0,0,0,0)$, the generators $\ttb^\mu$ are space-like due to \eqref{eq:orthogonalofPY}, and therefore the $t^i$ can be used to parametrize the internal 2-sphere $S^2_\tJ$. 
Even though that sphere has symplectic measure $\varpi$ \eqref{S2-volume-form} with volume $\tJ$
\begin{align}
\label{S2-volume-form-2}
    \int_{S^2} \varpi=\tJ\,.
\end{align}
reminiscent of a fuzzy sphere and thus admitting only  spherical harmonics $\hat Y(u)$ truncated\footnote{The truncation 
of spin can be understood in terms of the angular momentum operators $\m^{ij}$ acting on the internal $\cH_\tJ \cong \C^{\tJ+1}$ at the given reference point, see \cite{Sperling:2018xrm} for a detailed discussion. This $\cH_\tJ$ cannot be separated from the underlying $\cH$ globally, reflecting the Poisson structure on $\P^{1,2}$ which is dominated by $\{y^i,\ttb_j\}$. A careful computation of the symplectic form on $\P^{1,2}$ \cite{broukal} reproduces indeed \eqref{S2-volume-form-2}, confirming the cutoff $\tJ$ on $S^2_\tJ$.}  
 at spin $\tJ$, the $\ttb^i$ are {\em very different} from standard fuzzy sphere generators. 
In particular, 
\begin{align}
    \{\ttb^\mu,\ttb^\nu\} = -\frac{\theta^{\mu\nu}}{\ell_p^2 R^2}  = \frac{1}{R^2} m^{\mu\nu}, 
    \qquad \{u^{\mu},u^{\nu}\} =-\frac{1}{R^2\cosh^2(\tau)}\theta^{\mu\nu}= \frac{4}{\tJ^2\cosh^2(\tau)}m^{\mu\nu}
\end{align}
are {\em not} the inverse of the symplectic form
$\varpi$
but are suppressed by a large factor $\cosh^2(\tau)$. This implies that for late times $\cosh(\tau)\gg 1$,  the $\ttb^\mu$ 
are effectively commutative.
For example,
 the Poisson brackets
\begin{align}
    \{u^i,\{u^j,u^k\}\}
    = \frac{4}{\tJ^2\cosh^2(\tau)}\{u^i,m^{jk}\}
     \approx \frac{4}{\tJ^2\cosh^2(\tau)}(-\delta^{ij}u^k + \delta^{ik}u^j)
\end{align}
coincide with those on a fuzzy sphere with radius 1, but {\em  are suppressed by a factor $\cosh^2(\tau)$}.
Equivalently, the brackets of $\ttb^\mu$ coincide with those of a fuzzy sphere $S^2_\tJ$ of radius $\frac 1{\ell_p}$, but describe a sphere with radius $\frac 1{\ell_p}\cosh(\tau)$   \eqref{S2sphereM13}.
Hence the $\ttb^i \sim \sinh(\tau) p^i +  \tilde \ttb^i$ should be thought of as (compactified) momentum generators on $\cM^{1,3}$ with a small admixture fuzzy sphere generators $\tilde \ttb^i$.
This justifies the simplification $[\ttb \l,.] \to \ttb[\l,.]$ in the main text.

\subsection{Some loop integrals}\label{sec:loop-int}
We need the following  integral arising in the 1-loop effective action 
\begin{align}
\label{trace-EndM-kappa}
     \Tr_{\Mat(\cH_{\cM})}\Big( e^{-\im\a(\Box_{1,3}-\im\varepsilon)} f(y) \Big)
      &= \frac{1}{(2\pi)^4} \int_{\cM^{1,3}} d^4y\sqrt{G}f(y)
      \int\frac{d^4 k}{\sqrt{G}}
     e^{-\im\a k \cdot k}   \nn\\
     &= \frac{1}{(2\pi)^4} \int_{\cM^{1,3}} d^4y\sqrt{G}\rho^{-4} f(y)
      \int \frac{d^4 k}{\sqrt{G}}
      e^{-\im\a k_\mu k_\nu G^{\mu\nu}}  \nn\\
      &= -\im \frac{\kappa[\a]}{(2\pi)^4} \int_{\cM^{1,3}}d^4y\sqrt{G} \rho^{-4} f(y) 
\end{align}
for some function $f(y)$ on $\cM^{1,3}$,
where $k \cdot k = \gamma^{\mu\nu} k_\mu k_\nu$,  and we define
\begin{align}\label{eq:kappa-a}
    \kappa[\a] :&= \im \int\frac{d^4 k}{\sqrt{G}}
     e^{-\im\a k_\mu k_\nu G^{\mu\nu}} 
    =  \im\int d^4 k  e^{-\im\a k_\mu k_\nu \eta^{\mu\nu}} \nn\\
    &=  -  \int d^4 k_E\, e^{-\im\a k_\mu k_\nu \delta^{\mu\nu}} 
    = - 2\pi^2 \int_0^\infty dk_E k_E^3
   e^{-\im\a(k_E^2)}
   = \frac {\pi^2}{\a^2} .
\end{align}
by integrating first over $k^0$
via a contour rotation $\int d k^0 = \im \int d k^4_E$.



\setstretch{0.8}
\footnotesize
\bibliography{twistor}
\bibliographystyle{JHEP-2}

\end{document}